\documentclass[ 
reprint,
superscriptaddress,
 aps,prd,
 colorlinks=true, citecolor=green!50!black, linkcolor=persianblue!70!black, urlcolor=green!50!black, hypertexnames=false
]{revtex4-2}

\usepackage{amsmath}
\usepackage{amssymb}

\usepackage{subfigure}

\usepackage{amsxtra}
\usepackage{mathrsfs}
\usepackage{amsfonts}
\usepackage{arydshln}
\usepackage{dsfont}
\usepackage{dcolumn}
\usepackage{bm}
\usepackage{graphicx}
\usepackage{epstopdf}
\usepackage{txfonts}
\usepackage{braket}
\usepackage{comment}
\usepackage{blkarray}
\usepackage{xcolor}
\usepackage{color, colortbl}
\usepackage[normalem]{ulem}

\usepackage{tikz}
\usepackage{tcolorbox}

\usepackage{color}
\usepackage{subfigure}

\usepackage{orcidlink}

\usepackage{booktabs} 
\usepackage{siunitx} 
\usepackage{threeparttable} 
\usepackage{tabularx}
\usepackage{multirow}
\usepackage{diagbox}

\definecolor{niceblue}{rgb}{0.388235, 0.627451, 0.847059}
\definecolor{nicered}{rgb}{0.7,0.1,0.1}
\definecolor{nicegreen}{rgb}{0.1,0.5,0.1}
\definecolor{darkmagenta}{rgb}{0.55, 0, 0.55} 
\definecolor{persianblue}{rgb}{0.11, 0.22, 0.73}
\definecolor{LightCyan}{rgb}{0.88,1,1}

\usepackage{hyperref}

\usepackage{atbegshi,picture}
\AtBeginShipoutNext{\AtBeginShipoutUpperLeft{%
  \put(\dimexpr\paperwidth-1cm\relax,-1.5cm){\makebox[0pt][r]{\footnotesize \texttt{CETUP-2023-023   FERMILAB-PUB-24-0284-T}}}%
}}

\usepackage{aas_macros}

\begin{document}

\preprint{...}

\title{Does the Sun have a Dark Disk?}

\newcommand{\fnal}{\affiliation{Fermi National Accelerator Laboratory, 
Batavia, IL 60510, USA}}

\newcommand{\USP}{\affiliation{%
Departamento de F\'{\i}sica Matem\'atica, Instituto de F\'{\i}sica Universidade de S\~ao Paulo, 05315-970 S\~ao Paulo, Brasil
}}
\newcommand{\ukyphys}{\affiliation{Department of Physics and Astronomy, University of Kentucky, Lexington, KY, 40506-0055, USA}}

\author{Gustavo F. S. Alves\,\orcidlink{0000-0002-8269-5365}}
\fnal
\USP

\author{Susan Gardner\,\orcidlink{0000-0002-6166-5546}}
\email{susan.gardner@uky.edu}
\ukyphys

\author{Pedro~Machado\,\orcidlink{0000-0002-9118-7354}}
\fnal

\author{Mohammadreza Zakeri\,\orcidlink{0000-0002-6510-5343}}
\ukyphys


\begin{abstract}
The Sun is not quite a perfect sphere, and its oblateness, 
thought to be 
induced
through its rotation, 
has been measured using optical observations of its 
radius. Its gravitational quadrupole moment can then be 
deduced 
using
solar models, or through helioseismology, and it 
can also be determined from measurements of 
its gravitational effects on 
Mercury's orbit. 
The various assessments
do not appear to agree, with the most complete and precise orbital assessments 
being in slight excess of other determinations. 
This may speak to the existence of 
a non-luminous disk or ring, where we also note evidence for a circumsolar
dust ring within Mercury's orbit from the Solar TErrestrial RElations Observatory (STEREO) mission.
Historically, too, 
a protoplanetary disk may have been 
key to reconciling the Sun's metallicity 
with its neutrino yield. 
 The distribution of the non-luminous mass within Mercury's orbit can modify the relative size of the 
optical and orbital quadrupole moments in 
different ways. 
We develop how we can use these findings to 
limit the mass of a dark disk, ring, or halo in the immediate vicinity of 
the Sun, and we note how future observational studies of the
inner solar system
can not only refine these constraints but also 
help to identify and to assess the mass of 
its
dark-matter component.

\end{abstract}

\maketitle

\section{Introduction}
\label{sec:intro} 
As first noted by Dicke in 1964, an optical measurement of the solar oblateness of sufficient size 
could contribute significantly 
to Mercury's
perihelion precession and thus test Einstein's theory of general 
relativity (GR)~\cite{Dicke1964sunrelativity}~\footnote{
Dicke~\cite{Dicke1964sunrelativity}
noted that 
the existing GR test from the measured excess of 
Mercury's perihelion precession, of claimed 
$\sim \!1\%$ precision, neglected 
the effects of solar oblateness. Its inclusion 
could thus open the possibility of non-GR contributions.
}. 
A subsequent measurement of 
the oblateness $\Delta_{\odot}$, defined as the difference in the equatorial 
and polar radii over the mean solar radius, gave $(5.0 \pm 0.7)\times 10^{-5}$, apparently 
challenging GR at the sub--10\% level~\cite{Dicke1967SolarOblat}. 
In the intervening decades, 
the ability to assess the oblateness
has shown steady and significant progress, and space-based
studies have also been made, yielding refined errors. 
Moreover, the impact of magnetic-field-correlated
brightness variations on optical measurements
of the oblateness have been noted and
quantified~\cite{Rozelot2011historyoblateness}, 
removing disagreements and 
mitigating earlier puzzles~\cite{Dicke1967SolarOblat,DickeKuhnLibbrecht1986OblatMeasurement,Dicke1987J2Variable}. 
Ultimately, with improved 
measurements and theory, the gravitational quadrupole moment $J_2$ of the Sun 
can be determined, with a nonzero value
of about $2\times 10^{-7}$ --- and refined determinations of
its value from improved measurements of Mercury's orbit also 
support its approximate value, with the 
determined errors 
being some $100-1000$ 
times smaller than that, 
as we shall detail. 
These new levels of sensitivity open a new frontier, in that once negligible effects,
such as a circumsolar mass, 
can become appreciable. 
Here 
we compare the visual and
gravitational assessments of $J_2$ to 
constrain the mass and distribution
of non-luminous matter in 
the immediate vicinity of the Sun.

Different lines of evidence, from different
eras in the solar system's history, 
point to the existence of a non-luminous disk. 
For example, 
the measured energy spectrum of the solar 
neutrino flux determines 
the strength of the CNO 
cycle in the sun~\cite{BOREXINO2020CNOdiscovery,BOREXINO2022CNO,BOREXINO2023CNO}
with a rate compatible with 
high metallicity~\cite{Magg2022highmetalSun} 
but not 
low metallicity~\cite{Asplund2021lowmetalSun}
solar models. 
Concomitantly, there is also a long-standing
inconsistency between the element abundances
determined from the spectroscopy of the 
surface, as in \cite{Asplund2021lowmetalSun}, 
and those inferred from the interior
through helioseismology~\cite{Basu2008solarmodelprob,Zhang2019solarmodelling,Asplund2021lowmetalSun}.
This solar modelling problem can be mitigated 
if the interior metallicity of the  
 Sun can differ from that of its surface~\footnote{As noted
by \cite{Haxton2008CNCycle}, such would be at odds with
the standard assumption of a homogeneous
zero-age Sun.}, possibly through 
the formation of the photosphere with 
the gas-giant planets in the early solar 
system~\cite{Haxton2008CNCycle}. 
Later work has shown 
a metallicity gradient can appear 
if the sun formed while within a
circumsolar disk~\cite{Kunitomo2021imprint,
Kunitomo2022solarnudisk}.
Evidence has also been found 
for a circumsolar dust ring 
at the approximate location of 
Mercury's orbit~\cite{Stenborg2018dustring}.
The excess mass density is estimated at 5\%, and the 
total mass is not determined~\footnote{R.~Howard and 
G.~Stenborg, private communications.}.  
The origins of such a dust ring are not known,
and it could also stem from effects later in the
solar system's history~\cite{Pokorny2023Mercuryimpact}.
A non-luminous circumsolar disk could also 
contain a non-Standard Model, or dark matter,  
component. 
The capture of such an exotic component could
be considerably enhanced if ordinary matter in a 
non-luminous disk exists over a significant period
in the solar system's history. 
We note \cite{Brecher1979ring,Rawal2012rings,Iorio2012rings} for earlier 
discussion of the possibility of
massive circumsolar rings within 
Mercury's orbit
--- and possibly of dark matter~\cite{Iorio2012rings}.

We conclude this section with a sketch 
of the balance of the paper. The comparison of
different assessments of the solar quadrupole
moment is key to our ability to probe 
the matter distribution, and indeed dark matter, 
in the inner solar system. Thus we discuss the
various methods in some detail. We first
describe how the observed solar oblateness
connects to $J_2^{\rm Opt}$, the visible
quadrupole moment, before describing
how Mercury's measured perihelion 
precession, along with other data, can 
be used to determine $J_2^{\rm Orb}$, the
quadrupole moment determined through
gravitational interactions. Here we note
that, through use of helioseismology, that
$J_2^{\rm Heli}$ can also be found, through different methods,
though optical measurements of the Sun's surface are also
employed.
We then discuss possible non-luminous
components, including dark matter, and 
their
origin, that could exist within Mercury's orbit. We emphasize that 
the different possibilities influence
the relative sizes of $J_2^{\rm Opt}$, 
$J_2^{\rm Heli}$, and $J_2^{\rm Orb}$
differently. After considering 
the world's
data on $J_2$, we describe how the 
pattern of existing results can limit
the mass and distribution of non-luminous matter. 
Although the existing $J_2$ results may be impacted by 
unassessed  
systematic errors, we find the prospect of physical 
differences in their assessment to be an 
intriguing idea worthy of exploration.
Finally we describe future prospects of 
observational studies of the inner solar
system in making 
our final summary. 
In so doing we offer a perspective 
on the possible evolution of non-luminous
matter constraints, 
on their possible impact on future, refined GR 
tests~\cite{Will2018newGRperihelion}, 
and ultimately 
on our ability to distinguish conventional non-luminous matter from (exotic) dark 
matter in the inner solar system.
Although we have emphasized various 
determinations of the 
gravitational quadrupole moment, 
these studies naturally yield limits 
on solar mass loss, which could also have a dark-matter component, and on 
odd gravitational moments as well, and we note 
these and their implications as they arise.

\section{Oblateness and 
the quadrupole moment}
\label{sec:oblat_J2}
We first develop how observations of the Sun's surface can be
used to determine its oblateness and finally its quadrupole moment. 
A static, isolated, spherical Sun is perturbed by its rotation. 
If its center of mass (CM) is at rest and centered 
at the origin of a parametrized post-Newtonian (PPN) coordinate system, 
its gravitational potential, external to its surface ($r > R_{\odot}$), 
becomes~\cite{Misner1973Gravitation,Mecheri2004J2helio} 
\begin{equation}
    \phi_{\rm o}(r, \theta) = -\frac{G M_{\odot}}{r} \left[1 - \sum_{2n} \left(\frac{R_{\odot}}{r}\right)^{2n}\, J_{2n}\, P_{2n}(\cos \theta) \right] \,, 
    \label{eq:gravphio}
\end{equation}
where $\theta$ is the polar angle from the symmetry axis (co-latitude) and the 
$P_{2n}$ are Legendre polynomials. 
The solar mass $M_{\odot}$ is not
absolutely known, nor is $J_2$, the quadrupole-moment parameter, in that 
from the perspective of Mercury's orbit
they are
effective quantities that can be modified by 
mass in the Sun's immediate vicinity.  
In so noting we suppose any such excess mass to share its CM with that of the Sun. 
We also note that the apparent
azimuthal symmetry of Eq.~(\ref{eq:gravphio}) is not a limitation, because  
the solar studies we employ are nearly continuous through the Sun's rotational period, making the $J_{2n}$ azimuthally averaged quantities.
Moreover, if 
either external forces act on the Sun's CM
in an appreciable way
or if non-luminous matter
within Mercury's orbit 
shifts the Sun's CM from its 
visible one, 
then the gravitational potential need no longer 
be reflection symmetric with respect to its mid-plane 
and terms odd in $n$ can appear, via $2n \to n$ in Eq.~(\ref{eq:gravphio}). 
(We note \cite{Gardner:2020jsf_torque,Hinkel:2020tii_torque_data} finds a non-zero  torque on stars within some $3\,\rm kpc$ of the Sun.)
Introducing the effective potential $\Phi = \phi_{\rm o} - \phi_{\Omega}$
at the surface, where $\phi_{\Omega}$ is 
generated by rotation, the oblateness $\Delta_{\odot}$
can be connected to the gravitational moments
via the equipotential condition 
$\Phi (R_e) = \Phi (R_p) $ to find, 
working through $n=4$, 
\begin{equation}
    \Delta_{\odot} \equiv \frac{R_e - R_p}{R_{\odot}} \approx J_1 + \frac{3}{2} J_2 + J_3 + \frac{5}{8} J_4 + \frac{\Omega^2\, R_{\odot}^3}{2 G M_{\odot}}\,,
    \label{eq:DickeJ2plus}
\end{equation}
where $R_e$, $R_p$, and $R_{\odot}$ are the equatorial, polar, and mean solar radius, respectively --- and an effective rotation rate $\Omega$ emerges under
the assumption of axially symmetric differential rotation. We refer to 
Appendix~\ref{sec:app:oblateness} for all details. 
Neglecting all $J_i$ save for $J_2$ 
yields the long-known result of \cite{Dicke1970oblatenessJ2ApJ}, whereas 
setting $J_1=J_3=0$ yields that of \cite{Mecheri2004J2helio}.
If $J_2$ is determined from $\Delta_{\odot}$ with $J_{i\ne 2} =0$, we term
it $J_2^{\rm opt}$. 

\section{Interpreting the perihelion precession}
\label{sec:interp_peri_preces}
Some 92\% of Mercury's perihelion precession stems
from precisely known perturbations from the planets,  
particularly 
from Venus, 
Jupiter, and the Earth/Moon system~\cite{Roy2005ormo.book,Park2017MESSENGER}. 
Effects from non-Newtonian
gravitational effects and the Sun's $J_2$ 
are also well-known 
but are 
more poorly determined~\cite{Roy2005ormo.book,Park2017MESSENGER} --- and their 
most complete and precise determinations come from  analyses~\cite{Park2017MESSENGER,Genova2018MESSENGER}
of near-Mercury 
radio ranging and Doppler tracking data from the NASA 
MErcury Surface, Space ENvironment, GEochemistry, and Ranging (MESSENGER) spacecraft mission~\cite{Solomon2001MESSENGERmission}. 
Here Mercury's mean orbit frame is defined with respect to the International 
Celestial Reference Frame~\cite{Ma2009ICRF}, whose variation in orientation is sufficiently 
small that its impact on the perihelion determination lies beyond the 
sensitivity of  
the MESSENGER data~\cite{Park2017MESSENGER}. 
With this choice the perihelion 
precession rate is ultimately determined from the slope of the 
phase angle determined
from fitting the data with a function containing both 
steadily increasing and periodic time-dependent features. The quantity 
$\dot{\varpi}$,
the rate of perihelion precession along Mercury's orbit plane, appears linearly in time $t$ and 
contains the gravitoelectric (GE), Einstein-Lense-Thirring (ELT), 
and 
solar quadrupole moment 
contributions within it. 
In the PPN formalism, the GE effect is linear
in $\beta$ and $\gamma$ (and is nonzero in GR~\cite{Einstein1916AnP...354..769E}, 
in which $\beta=\gamma=1$), with the ELT effect being numerically nearly
negligible 
if the solar angular momentum 
$S_{\odot}$ determined from helioseismology 
is used~\cite{Pijpers1998helioJ2}. Physically, we note $\beta$ and 
$\gamma$ describe the nonlinearity in the superposition law for 
gravitation and the space curvature produced by a unit rest mass, respectively~\cite{Misner1973Gravitation}. 
In \cite{Park2017MESSENGER}, e.g., the ELT effect is absorbed within their error analysis. 
Nevertheless, a determination of $\dot{\varpi}$ contains
$\beta$, $\gamma$, and $J_2$, so that 
additional information is needed to separate them. Using 
information on $\gamma$ from
the Cassini mission, $\gamma-1 = (2.1\pm 2.3)\times 10^{-5}$~\cite{Bertotti2003Natur_Cassini} 
and the time structure of the precession angle, 
both 
$\beta$ and $J_2$
can be determined~\cite{Park2017MESSENGER}. 
The analysis of 
\cite{Park2017MESSENGER} uses MESSENGER data from its four-year orbital phase (2011-2015), whereas that of \cite{Genova2018MESSENGER} uses all of the 
MESSENGER data over 
its fly-by and orbital 
phases (2008-2015). These analyses 
can employ the Nordtvedt parameter~\cite{nordtvedt1968equivalence, 1972ApJ_Will_Nordtvedt} constraint 
$\eta=4\beta-\gamma-3=0$ of GR as well. 
The values of $J_2$ from these studies and others 
are collected in the Tables in Appendix \ref{sec:app:tables};
we will also consider the 
values that emerge if $\beta=\gamma=1$, as in GR. 
These studies also offer constraints on time-dependent effects, particularly 
through the apparent time-independence of the residuals in the 
perihelion precession fit of \cite{Park2017MESSENGER} and 
of the estimated rate of change in the solar gravitational 
parameter $\mu\equiv G M_{\odot}$. 
Namely,
$\dot{\mu}/\mu 
= (-6.13 \pm 1.47)\times 10^{-14}\,\rm yr^{-1}$~\cite{Genova2018MESSENGER}, to be compared
with the planetary ephemerides result of 
$(-10.2\pm 1.4) \times 10^{-14}\, \rm yr^{-1}$~\cite{Pitjeva2021masschange}, 
where we note 
\cite{Fienga2024review} for a review. 
We return to the implications of these results in Sec.~\ref{sec:lim:non-lum}. 

\section{Possible non-luminous components}
\label{sec:non-lum}

Although the dust in the solar system can reflect light~\cite{Leinert1981dust}, 
it does not emit visible light, 
and thus we have classified it as non-luminous matter. 
A dusty disk appears to exist throughout much of the solar system~\cite{Leinert1980dustplane}, 
and simulations suggest that its structure 
differentiates between 
the inner and outer solar system~\cite{Brasser2020protodisk}.
A population of micrometeroids, seeded by collisional grinding of zodiacal dust, may also be relevant within 1 AU~\cite{Sommer2023micrometeroid} and
still other sorts of nonluminous objects may contribute~\cite{Taylor2024darkcomet}.
We have noted 
that a ring of dust has been discovered in the path of Mercury's 
orbit~\cite{Stenborg2018dustring}, from the STEREO mission~\cite{kaiser2008stereo}, and a model of its mass distribution suggests its 
mass could be about 
$(1.02 - 4.05) \times 10^{12\pm (\approx 1)} \,\rm kg$, roughly equivalent to the mass of a single asteroid~\cite{Pokorny2023Mercuryimpact}. Known asteroids
range from about $10^{10-21} \, \rm kg$ in mass~\cite{asteroiddata}.

Yet dust is not the only possibility. Dark matter, e.g., may also contribute. 
Studies of stellar tracers in the solar neighborhood 
suggests that the local dark matter density is only some 
$\rho_{\rm dm} = 0.4\,\rm GeV/cm^3$~\cite{Gardner2021review}, and 
a spheroid of that density 
would contribute
a mass of 
$6\times 10^{11} \, \rm kg$, 
or $3\times 10^{-19} \rm M_\odot$, 
within Mercury's orbit. 
Moreover, it has long been thought that gravitational focusing 
mechanisms exist, modifying the dark-matter
velocity distribution and acting 
to increase the 
dark matter density within 
the solar system~\cite{Sikivie2002solarwakesDMflow,
Alenazi2006DMunboundSun,Peter2009DMSolarSys,Khriplovich2009DMcapture, 
Lee2013gravfocusingdirectdet,Sofue2020gravfocusing,Kim2022gravfocusingwaveDM},
albeit the total mass added is not very significant.
The detailed estimates depend on the dark-matter 
model and on astrophysical modelling. For example, upon 
adopting a WIMP dark-matter candidate within a standard Galactic 
halo model, the direct detection event rate 
for gravitationally captured WIMPs on the solar system has been
determined to never exceed 0.1\% of the event rate for halo 
WIMPs~\cite{Peter2009DMSolarSys,Peter2009DMSolarSysIII}, 
though 
gravitational perturbations external to the solar system could 
impact the bound WIMP population significantly~\cite{Peter2009DMSolarSysIII}. 
More significant
variations should be possible in models for which the dark-matter candidate
possesses inelastic interactions, as constraints from Liouville's theorem
would not apply. Gravitational focusing of unbound
dark matter on individual planetary bodies has also been found to 
yield significant, local density enhancements in their immediate 
vicinity~\cite{Alenazi2006DMunboundSun,Sofue2020gravfocusing}
and 
even ring-like, time-dependent effects~\cite{Alenazi2006DMunboundSun} 
that we shall investigate further.

There have also been studies of the
capture of dark matter within 
celestial bodies, and for models
with sufficient  
dark-matter--matter 
interactions, significant enhancements over the nominal dark-matter density are possible~\cite{Leane2022floatDM,Leane2023DMcapture,Dasgupta2019DMcapture,Dasgupta2020DMcapture,Bramante2023LDMcaptureplanets,Ray2023strongintDM}, 
showing that 
dark matter could 
contribute non-trivially 
to the mass of a celestial
body within its observed 
radius. Measurements of the Earth's heat budget do 
limit, however, the possibility of strong  dark-matter-matter
interactions~\cite{Mack:2007DMearthheat,Adler:2008rq}, though some of the internal 
heating of Jovian planets could speak to planet-bound
dark matter~\cite{Adler:2008ky}. 

A celestial body can also possess a dark halo, and this can emerge
if the dark-matter particle simply possesses self-interactions. 
In the ``gravi-atom'' 
mechanism of \cite{Budker2023gravatom}, for example, 
two-body scattering, particularly of ultra-light dark matter, 
in the 
gravitational field 
of a celestial body 
can yield dark-matter capture 
and ultimately a dark halo. We 
will 
consider this scenario further in 
Sec.~\ref{sec:dm_impl}. If the dark matter couples to neutrinos, then 
neutrino oscillation data can probe the Earth's dark halo~\cite{Gherghetta:2023myo}. 

The possibility of exotic dark structures, such as exoplanets~\cite{Bai2023darkexoplanet}, 
filaments~\cite{Prezeau2015denseDMhairs}, or other macroscopic objects~\cite{Korwar2023macroDM}, 
have also been  suggested. 
These macroscopic objects 
need not have a dark matter
component, as, 
say, in the case of magnetic black holes~\cite{Bai:2020magneticBH}, but others, such as 
primordial black holes (PBHs), can also 
function as a dark matter candidate and 
are 
as yet poorly constrained in the asteroid mass
range~\cite{Green2024PBHdarkmatterrev}. Their transits
of the solar system can 
be limited through 
planetary 
ephemerides~\cite{Tran2023PBHdarksolarsystem}, as well
as via observational means~\cite{Ray2021PBH_MeVtelescope}. Gravitational wave detection can also probe the possibility of local,  macroscopic dark matter~\cite{Du:2023MacroDM_GW} and of the solar system's mass distribution~\cite{Takahashi2023ApJgravsolar}. 
The Sun and 
planets can also emit light
dark particles, as, e.g., in 
\cite{Sikivie1983axionhelio,Feng2016DarkSunshine}. However, 
the time rate of change in the gravitational parameter is smaller than that estimated from the net effect of dust capture 
and the solar wind~\cite{Genova2018MESSENGER}, so that 
no limit on dark-matter
emission is currently possible.

Constraints on all these
scenarios emerge from 
measurements of planetary motion \cite{Anderson:1988BoundsDMSolar,Anderson:1995BoundsNonluminousMatterSun,Gron:1995rn, Sereno:2006mw,
Khriplovich2006dmsolarsys,
Frere:2007pi,
Pitjev2013constraints}, which 
 show that the maximum density at the Earth's orbit is limited to be below $10^2-10^5$ GeV/cm$^3$, depending slightly on the profile shape. 
Somewhat weaker constraints
emerge from studies 
of the inner planets~\cite{Pitjev2013constraints}. 
We emphasize that existing 
estimates of 
dark-matter capture 
on the solar system, including work within the popular ``WIMP'' dark-matter
paradigm, 
fall rather short of these limits~\cite{Peter2009DMSolarSys,Khriplovich2009DMcapture}. 
There are also constraints on the Earth's own dark halo~\cite{Adler:2008ky}, and
on the dark matter density by analyzing the propagation of light in the solar system~\cite{Arakida:2008sh}. 
Dark matter would cause time delay and frequency shift of the 
light, though the 
constraints are not very stringent~\cite{Arakida:2008sh,Gardner:2009cosmicn}. 

In this paper we develop constraints on 
the mass and distribution 
of non-luminous matter external to the Sun and 
planets 
and thus of
dark-matter models connected
to these possibilities. The suggested 
magnitude of the excess mass we find, were it to come from 
exotic sources, points preferentially to the possibility of  
non-WIMP dark-matter candidates, as we detail in Sec.~\ref{sec:dm_impl}.

\section{Gravitational Quadrupole Moment Determinations} 
\label{sec:gravJ2}
Determinations of $J_2$ began decades ago and also range over decades.
Three distinct methods have been used. 
There have been direct optical measurements of the 
solar oblateness $\Delta_\odot$, which, when combined with theory, as in 
Eq.~(\ref{eq:DickeJ2plus}), yield $J_2^{\rm Opt}$. Helioseismological data probes
the structure of the Sun; this, when combined with a model for its structure, 
can be used to infer $J_2^{\rm Heli}$. Finally, as we have noted, precision measurements of planetary
motion 
can also be used to infer $J_2^{\rm Orb}$. We emphasize
that the orbital determinations are sensitive to the existence of 
mass {\it within} the orbit in question. 
In this paper we focus on determinations
of $J_2^{\rm Orb}$ from measurements of Mercury's orbit, to address
the possibility of non-luminous matter 
in the immediate region of the Sun. 
There is a long history of solar oblateness
measurements~\cite{Dicke1967SolarOblat,DickeKuhnLibbrecht1986OblatMeasurement}, with improved assessments once space-based studies became 
possible~\cite{Kuhn2004ConstancySunDiameter}. Surface magnetism and
other effects can impact the observed shape~\cite{Emilio2007ChangingSolarShape,Fivian2008SolarOblatMag}, and 
Fivian \emph{et al.} have provided a corrected assessment that would
remove the enhancement from solar magnetism~\cite{Fivian2008SolarOblatMag}.
However, we note that 
discussions of the shape assessment, and the possibility of a time-dependent solar shape, continues~\cite{Kuhn2012precisesolshape,basu2019solarrotationchange,Eren2023temporalsolarJ2}.

We summarize the most 
pertinent 
determinations here and note 
Appendix~\ref{sec:app:tables}
for a more complete list. 
We report
the results for $J_{n}$ in units of $10^{-7}$ throughout. 
From Park \emph{et al.}~\cite{Park2017MESSENGER} 
\begin{equation}
    J_2^{\rm Orb} = 2.25 \pm 0.09 \,\rm \,;\,
    J_2^{\rm Orb} |_{\beta=\gamma=1} 
    = 2.28 \pm 0.06  \,,
\end{equation}
whereas from Genova \emph{et al.}~\cite{Genova2018MESSENGER}
\begin{equation}
    J_2^{\rm Orb} = 2.246 \pm 0.022 \,\rm \,;\,
    J_2^{\rm Orb} |_{\beta=1; \eta=0} 
    = 2.2709 \pm 0.0044  \,.
\label{eq:J2Genova_Orb}    
\end{equation}
Here we quote the quadrupole moment assessment 
within a PPN framework, as well as its value upon assuming GR. In the first case, 
Genova \emph{et al.} 
also find the PPN parameter 
$\eta=(-6.6 \pm 7.2)\times 10^{-5}$, noting
that nonzero $\eta$ can be associated with a
shift in the solar system barycenter~\cite{Genova2018MESSENGER} --- and
vice versa. 
We note appreciable shifts in the central
values and smaller uncertainties are associated with the GR limit. 
Since studies of dark matter always assume that GR is valid, 
we choose $\beta=\gamma=1$ ($\eta=0$) for our studies here. Moreover, 
the result of Genova {\emph et al.}~\cite{Genova2018MESSENGER} reflects the use of the
full MESSENGER data set, so that 
it would seem we should
we 
employ the GR limit in 
Eq.~(\ref{eq:J2Genova_Orb}) for $J_2^{\rm Orb}$ henceforth.

Before proceeding, however, we pause to consider the gross discrepancy 
in the errors in the two determinations, as they are significantly different. 
The origin of this has been 
investigated by Konopliv, Park, and Ermakov~\cite{Konopliv2020Icar..33513386K}, 
and it appears that the difference largely stems from a simplifying assumption 
in the Genova et al.~\cite{Genova2018MESSENGER} analysis, namely, that the Earth's orbit is perfectly known. When the uncertainties in the orbit of the Earth and Mercury generated by all the other planetary bodies are included, then the error in 
$J_2^{\rm Orb}$ increases by about a factor of 4,  
explaining the difference in the errors found in the two analyses 
in the PPN case~\cite{Konopliv2020Icar..33513386K}. Presumably  
a similar enlargement appears in the GR case as well, though it would
not explain the difference in the errors in the two $J_2^{\rm Orb}$ 
determinations. In what follows, we will employ the GR result of 
Eq.~(\ref{eq:J2Genova_Orb}), but be mindful of the expected larger error~\cite{Konopliv2020Icar..33513386K} in considering its significance. 

Turning to 
the optical assessment of 
$\Delta r \equiv R_e -R_p =(8.01 \pm 0.14)\, \rm milliarcsec\, (mas)$
due to
Fivian \emph{et al.}~\cite{Fivian2008SolarOblatMag} using the space-based 
Reuven Ramaty High Energy Solar Spectroscopic Imager
 (RHESSI) instrument~\footnote{We note 
\url{https://hesperia.gsfc.nasa.gov/rhessi3/} for further information and 
Ref.~\cite{Kuhn2012precisesolshape} for discussion of this result and their finding
of $\Delta r = 7.20 \pm 0.49\, \rm, mas$.}, 
the value of $J_2$ can be determined from
$J_2=(2/3)(\Delta r - \Delta r_{\rm surf})/R_\odot$~\cite{DickeKuhnLibbrecht1986OblatMeasurement} with
 $\Delta r_{\rm surf} \approx 
7.8\, \rm mas$~\cite{Dicke1970oblatenessJ2ApJ}
and $R_\odot = 9.5963 \times 10^{5} \,\rm mas$~\cite{Auwers1891solarradius}, its radius at $1\,\rm AU$~\footnote{This value is employed by Fivian 
\emph{et al.}~\cite{Fivian2008SolarOblatMag}, though it has been officially updated to $9.5923 \times 10^{5}\,\rm mas$ by the International Astronomical Union 
in 2015~\cite{Prsa2016IAUresolutionsolar}, with additional space-based measurements  
from PICARD/SODISM supporting this later result~\cite{Meftah2018solarradius}. For our purposes $J_2$ is sufficiently small that this refinement does not matter.}.
Noting $\Delta {r_{\rm surf}}/R_{\odot}$ corresponds
to the last term in 
Eq.~(\ref{eq:DickeJ2plus}), 
they find~\cite{Fivian2008SolarOblatMag}
\begin{equation}
     J_2^{\rm Opt} = 1.46 \pm 1.0 \,,
\label{eq:J2Opt}     
\end{equation}
where the error does not include an error in $\Delta r_{\rm surf}$. 
If we were, rather, to employ the $\Delta r$ measurement in \cite{Kuhn2012precisesolshape}, $\Delta r =7.20 \pm 0.49\, \rm mas$, the 
associated $J_2$ would be negative but consistent with zero with a much larger error. 
For clarity we reiterate that in 
this paper we define 
the {\it oblateness} as 
$\Delta_\odot
\equiv \Delta r/R_\odot$,  
which evaluates to 
$(8.35 \pm 0.15)\times 10^{-6}$, because, 
in contrast, Fivian \emph{et al.}~\cite{Fivian2008SolarOblatMag}
term $\Delta r$ the oblateness.

We now turn to the assessments from 
helioseismology. The $J_2^{\rm Heli}$ 
comes from observations of the oscillations
in the Sun's surface interpreted within a 
solar model. Here we highlight
the work of Mecheri and Meftah for their use of multiple, space-based solar oscillation data sets and multiple solar evolution models, all within 
an integral equation 
approach that eliminates the need for a solar differential rotation model~\cite{Mecheri2021J2helio}. 
This is an update of 
Mecheri \emph{et al}. (2004) that 
uses data from 
the Michelson
Doppler Imager on the 
Solar and Heliospheric Observatory (SoHO/MDI) and
models for the solar differential rotation. Using
models (a) and (b), respectively, from \cite{Corbard2002SolarModel} 
they report~\cite{Mecheri2004J2helio}
\begin{eqnarray}
  J_2^{\rm Heli}= 2.201 \,&;&\,	
   J_4^{\rm Heli} = -5.601 \times 10^{-2} \,. \\
     J_2^{\rm Heli}= 2.198 \,&;&\,	
   J_4^{\rm Heli} = -4.805 \times 10^{-2} \,. 
   \label{eq:J2heli_mercheri2004}
\end{eqnarray}
In contrast, 
$J_2^{\rm Opt}$ in 
Eq.~(\ref{eq:J2Opt}) follows from setting $J_4$ to zero, 
so that $J_4$'s size indicates
the size of the theoretical systematic error in that procedure. The outcome of 
\cite{Mecheri2004J2helio}
is reported as
$J_2^{\rm Heli} = 2.20 \pm 0.03$ by \cite{Genova2018MESSENGER} and results from combining all the $J_2$'s in Table II of \cite{Mecheri2004J2helio} (including
a uniform solar rotation result)~\footnote{The error was assessed by accounting for their differences and multiplying by two. 
A. Genova, private communication.}. The update of the 2004 analysis by \cite{Mecheri2021J2helio} 
reports 
$J_{2n}$ with $n=1\dots 5$
using data from either
the Helioseismic and Magnetic Imager on the Solar Dynamics Observatory (SDO/HMI) or SoHO/MDI using 
the online compilation of 
\cite{heliodata} with the 
analysis of \cite{Larson2015helioanal,
Larson2018helioanal}, 
upon the use of 
either the
CESAM~\cite{CESAM2008} or ASTEC~\cite{ASTEC2008} theoretical 
(solar evolution) frameworks. 
With CESAM, they report~\cite{Mecheri2021J2helio}
\begin{eqnarray}
{\rm SDO/HMI} : 
   J_2^{\rm Heli}= 2.211 \,;\,	
   J_4^{\rm Heli} = -4.252 \times 10^{-2} \,; 
  \label{eq:J2_SDOHMI_CESAM} \\
   {\rm SoHO/MDI} : 
  J_2^{\rm Heli}= 2.204 \,;\,	
   J_4^{\rm Heli} = -4.064 \times 10^{-2} \,.
\end{eqnarray}
Using ASTEC, they find~\cite{Mecheri2021J2helio}
\begin{eqnarray}
{\rm SDO/HMI} : 
   J_2^{\rm Heli}= 2.216 \,;\,	
   J_4^{\rm Heli} = -4.256 \times 10^{-2} \,; 
   \label{eq:J2_SDOHMI_ASTEC}
   \\
   {\rm SoHO/MDI} : 
  J_2^{\rm Heli}= 2.208 \,;\,	
   J_4^{\rm Heli} = -4.069 \times 10^{-2} \,.
\end{eqnarray}
The data sets SDO/HMI and SoHO/MDI are
independent and correspond to observations from 2010 Apr -- 2020 July
and 1996 May -- 2008 March, respectively. The slight differences
in the reported outcomes of the two data sets could 
stem from a time dependence~\cite{Rozelot2020tdephelio,Mecheri2021J2helio}
or from an observational systematic error that differentiates between the two instruments --- or both. 
We note that the 
SDO/HMI and MESSENGER results 
are roughly contemporaneous. 
Computing the average and standard deviation of the results using 
one data set
or the other, we find 
\begin{eqnarray}
{\rm SDO/HMI} : \,
 && J_2^{\rm Heli}= 2.214 \pm 0.002 \,;\,	\label{eq:J2_finalHelio} \\
   &&J_4^{\rm Heli} = (-4.254  \pm 0.002)\times 10^{-2} \,; \nonumber  \\
   {\rm SoHO/MDI} : \, 
 &&J_2^{\rm Heli}= 2.206 \pm 0.002 \,;\,	 \label{eq:J2SoHO/MDI} \\ 
  && J_4^{\rm Heli} = (-4.067 \pm 0.003)\times 10^{-2} \,. \nonumber 
  \end{eqnarray}
Here the error stems
entirely from the use of different solar models. 
We note that the midpoints 
of the observing periods for the SDO/HMI and SoHO/MDI data sets are 
separated by 11 years, where $J_2$ changes by $|\Delta J_2^{\rm Heli}|
=0.008 \pm 0.002$, since the errors
are not independent. 
For reference, Antia \emph{et al.}~\cite{Antia2008Helio} 
use the SoHO/MDI 
data to find a $J_2$ of 
$2.220 \pm 0.009$, where the
error denotes the estimated 
time dependence~\cite{Antia2008Helio}.  Given this, we 
suppose that 
$|\Delta J_2^{\rm Heli}|$ could indeed stem from time dependence, 
rather than from an observational systematic error. We thus
employ $J_2^{\rm Heli}$ in
Eq.~(\ref{eq:J2_finalHelio}), using SDO/HMI data, 
as the helioseismological outcome to be compared with 
orbital determinations of $J_2$ from the MESSENGER data, because they are approximately contemporaneous. 
Using the $J_2^{\rm Orb}$ result of
\cite{Genova2018MESSENGER} that
assumes GR, we find 
a small, albeit 
significant, difference
in the assessment of 
$J_2$ from orbital and solar measurements: 
\begin{equation}
J_2^{\rm Orb}|_{\beta=1; \eta=0} - J_2^{\rm Heli}
= 0.057 \pm 0.006  \,\,  (\pm 0.020)\,,
\label{eq:J2diff}
\end{equation}
where the error in parentheses replaces the value of $\pm 0.006$ 
if the estimated error in Earth's orbit~\cite{Konopliv2020Icar..33513386K} is taken into account. 
Of course the error in 
$J_2^{\rm Heli}$ in Eq.~(\ref{eq:J2_finalHelio}) that appears in 
these differences come 
from the use of different 
solar models, 
a systematic error of theoretical origin that may be underestimated,
and
we can wonder about how else to 
estimate it. 
Alternatively, we estimate  the error in $J_2^{\rm Heli}$ 
by using all four values in 
Eqs.~(\ref{eq:J2_finalHelio}) and (\ref{eq:J2SoHO/MDI}) demanding that $\chi^2=1$ for a flat-line fit. This gives an
average $J_2^{\rm Heli}$ of $2.210 \pm 0.004$, so that 
Eq.~(\ref{eq:J2diff}) becomes
$0.061 \pm 0.009$. 
Separately, as we detail in 
Appendix \ref{sec:app:tables}, 
if we compute an average and
standard deviation of our collected
helioseismological results, 
we find 
\begin{equation}
   \langle     J_2^{\rm Heli} \rangle \Big|_{\rm all}  = 2.213 \pm 0.002 \,, 
   \label{eq:J2heli-all}
\end{equation}
which, interestingly, is also consistent with 
Eq.~(\ref{eq:J2_finalHelio}). 

Since the assessment of the solar quadrupole moment from helioseismology, $J_2^{\rm Heli}$, is key to our analysis, we pause to consider the parameters on 
which it depends. The solar models employed are calibrated to precise helioseismic data, including the Sun’s internal sound-speed profile, rotation, and composition. 
They appear to be 
robust: variations in the composition, the equation of state, or the internal rotation typically change $J_2$ by less than 1\%~\cite{Mecheri2004J2helio,Roxburgh2001heliomeasurement}.
Recent work confirms and tightens this assessment, with 
solar models differing in composition or opacity yielding $J_2$ values 
that vary by $\lesssim 0.2\%$~\cite{Mecheri2021J2helio}.
We note that the integral that yields $J_2$ has significant support 
in the radiative zone, where helioseismic constraints permit 
little variation~\cite{Mecheri2021J2helio}.
For instance, 
Mecheri and 
Mertah~\cite{Mecheri2021J2helio} 
found that including differential rotation in the convection zone shifts $J_2$ by only about 0.5\%, while changes to core conditions produce even smaller deviations. In contrast, the central discrepancy we identify between 
the orbital determinations 
(e.g., $J_2^{\rm orb} \simeq 2.27 - 2.28$~\cite{Park2017MESSENGER,Genova2018MESSENGER} in the GR limit) 
and the helioseismic value ($J_{2,\mathrm{helio}} \simeq 2.21$) 
is $\simeq 0.06$, which is several times larger than
the spread determined from the noted modeling uncertainties. 
Nevertheless, since 
the size of the observed discrepancy challenges explanation by modeling assumptions alone, this suggests  that further investigation of its possible limitations 
is both justified and necessary.

The significance of the result 
in Eq.~(\ref{eq:J2diff}) also relies, 
however, 
on assuming GR values for the PPN parameters
$\beta$ and $\eta$, although 
this is the usual procedure 
if the possibility of dark matter is considered. 
Finally, 
referring to the Tables
in Appendix \ref{sec:app:tables}, 
we observe that 
each of the modern $J_2^{\rm Heli}$ assessments (after 1990) 
are numerically smaller than 
those relying on MESSENGER data~\cite{Park2017MESSENGER,Genova2018MESSENGER}, even without
assuming GR, 
though 
not all are significantly so. 
The tabulated values of the 
optical $J_2$ computed from the observed
oblateness show that the errors
are much larger, so that a similar
comparison with the orbital results is not possible. We also  
show alternate orbital $J_2$ assessments in Appendix \ref{sec:app:tables}. Analyses
in which the ELT effect is 
taken into account and the various
PPN parameters are fitted are 
arguably better and give a 
plausibly more robust 
inference of $J_2$ than studies that do neither.
The three different methods we have considered to assess $J_2$ probe 
the Sun's shape, but they are also sensitive to different mass distributions. Their difference can limit the presence of additional matter within Mercury's orbit, as we discuss next.

\section{Limits on mass and distribution of non-luminous matter} 
\label{sec:lim:non-lum}
Disagreements in the assessment of 
nominally equal quantities, as we have noted for $J_2$,   
can signal the presence of non-luminous matter within Mercury's orbit. 
If that matter were a spherical dark matter halo, then we would expect
its presence 
to reduce $J_2^{\rm Orb}$ with respect to visible assessments, yielding 
\begin{equation}
J_2^{\rm Orb} < J_2^{\rm Opt} \,,\, J_2^{\rm Heli} \,. 
\label{eq:darkhalo}
\end{equation} 
On the other hand, if a dark disk 
co-planar with the planetary orbits were present, 
then we would expect, rather, 
\begin{equation}
J_2^{\rm Orb} > J_2^{\rm Opt} \,,\, J_2^{\rm Heli} \,. 
\label{eq:darkdisk}
\end{equation} 
We can also use the 
differences in the $J_2$ assessments to limit
the excess mass enclosed, as we determine later in this section. 
Observational limits 
can also be placed on the change in the Sun's mass with time, namely $\dot{M}_{\odot}$, but, as we have noted, the
current limit is less than
the changes expected from 
dust deposition or mass loss through the action of the solar 
wind~\cite{Genova2018MESSENGER}. 
Also 
we can limit the possibility of unexpected gravitational perturbations on Mercury's orbit. 
Finally, since we have a ``visible'' assessment of $J_2$ determined
in two different ways, we can also use Eq.~(\ref{eq:DickeJ2plus}) to back out
a limit on the odd $J_n$ moments. We
note that $J_{2n+1}$ can be a signed quantity.  

We now work through these points in reverse order. 
To determine a limit on the odd $J_n$ moments we combine 
$\Delta_{\odot}$ from Fivian \emph{et al.}~\cite{Fivian2008SolarOblatMag},
which is from 2008, with the roughly contemporaneous assessment using 
SoHO/MDI data 
from Mecheri and Meftah~\cite{Mecheri2021J2helio}, averaged over solar models, as reported in 
Eq.~(\ref{eq:J2SoHO/MDI}). 
With this, 
using Eq.~(\ref{eq:DickeJ2plus})
and $\Delta r_{\rm surf}/R_{\odot}$ for
its last term, as in \cite{Fivian2008SolarOblatMag},  we thus find 
\begin{equation}
    J_1 + J_3 \simeq J_1 = -1.1 \pm 1.5 \,.
\end{equation}
We note that $J_1$ can easily be larger than $J_3$ because it can be 
generated by either external forces on the Sun or by a shift of
the solar CM from its visible one. Although 
$J_1$ is consistent with 
zero, it is intriguing that its central value obeys 
$J_1 <0$, which is compatible with the Sun's location north 
of the Galactic mid-plane. We recall that the inverse-square-law
nature of gravitational forces ensures that the 
acceleration of a star is typically determined by the 
mass distribution at large scales, rather than by stars
in its immediate vicinity~\cite{BinneyTremaine2008gady.book}.

Now we turn to the question that motivates this paper: 
how the amount of non-luminous mass, 
supposing some 
distribution, is  limited
through 
the different determinations of $J_2$. The constraints we develop in this section 
concern limits on the approximate total mass within Mercury's orbit. 
For reasons of simplicity, we consider the 
possibility of a dark disk or ring 
in the
plane of Mercury's 
orbit
and a spherical dark halo 
all centered on the
Sun's CM
--- though some mixture of these, or
some more complicated shape, could also be possible. 
We have noted that dust studies point to the existence of a dusty ring coinciding 
with the average path of Mercury's orbit~\cite{Stenborg2018dustring}, and
our studies limit its possible mass, but not just. Since dust, or conventional matter more generally, 
could act as a substrate for the ultra-light-dark-matter capture mechanism described 
in \cite{Budker2023gravatom}, or that other dark-matter enhancement mechanisms could operate, 
so that dark matter, i.e., non-standard, non-luminous matter, 
could also contribute appreciably to the non-luminous mass within Mercury's orbit. 
We note that the most
precisely determined 
$J_2$ values, 
as in Eq.~(\ref{eq:J2diff}), 
may differ fairly significantly from zero, 
suggesting that 
Eq.~(\ref{eq:darkdisk}) holds
and a non-luminous 
disk is favored. 
Nevertheless, in what
follows, we suppose
either a disk/ring or 
spherical halo scenario
and limit the associated
mass with each. 

{\it Limiting the mass of a non-luminous disk or ring:}
We conceptualize this object as an axially symmetric circumsolar ring potentially composed of both dust and dark matter, centered about the CM of the visible Sun.
It is characterized by a mass \( M_r \), a uniform density \( \sigma \), an inner radius \( R_i \), an outer radius \( R_o \), and a height \( h \) along the $\hat{\mathbf{z}}$-axis, defined as perpendicular to the plane of Mercury's orbit, 
noting 
that the Sun's rotation axis, 
at epoch 1950.0,
is tilted about 
$ 3.38^\circ$
with respect to it~\cite{Rozelot2011historyoblateness}.
In what follows, though, we will set that tilt to zero for simplicity. To compute its gravitational quadrupole moment, we assume the ring is also rigid. In a Cartesian coordinate system, the moment of inertia tensor is diagonal, and~\cite{fetter2003theoretical}
\begin{equation}
    J_2^{\rm ring} = 
    \frac{1}{M_{\rm r} {\bar R}^2} (I_z - I_x)
\end{equation}
with \( \bar{R} \equiv ( R_i + R_o ) / 2 \), where the moment of inertia in the direction $s$ is given by 
\begin{equation}
    I_s = \int d^3 r\,\rho(\mathbf{r})
    \left[ 
    r^2 - (\mathbf{r}\cdot 
    \hat{\mathbf{e}}^{(s)})^2 
    \right] \,.
\end{equation}
More concretely,
\begin{eqnarray}
    I_z &=& \frac{M_{\rm r}} {2}  \left[R_o^2 + R_i^2 \right] \,,\\
    I_x &=&
   M_{\rm r} \left[\frac{1}{4}\left(R_o^2 + R_i^2\right) + \frac{h^2}{12} \right]  \,
\end{eqnarray}
and thus 
\begin{align}
    J_2^{\rm ring} = \frac{1}{\bar{R}^2} \left[ \frac{1}{4}  \left(R_o^2 + R_i^2 \right) -\frac{h^2}{12} \right]. \label{eq:ring:j2}
\end{align}
Combining the potential of the dust ring with the Sun's for \( r > R_o \),
Eq.~(\ref{eq:gravphio}), yields:
\begin{equation}
    \phi_o^{\rm tot} =
     \frac{M_{\odot} + M_{\rm r}}{r} -  \left(\frac{J_2^{\rm int} M_{\odot} R_{\odot}^2 + J_2^{\rm ring} M_{\rm r} \bar{R}^2}{r^3}\right) \left(\frac{3\cos^2\theta - 1}{2}\right), \label{eq:pot:ring_plus_sun}
\end{equation}
where $J_2^{\rm int}$ is the intrinsic solar quadrupole moment, which may be measured through a direct optical measurement or via helioseismology. Defining the fractional mass of the ring, \( \epsilon_{\rm r} \equiv M_{\rm r} / M_{\rm tot} \), with $M_{\rm tot} \equiv M_{\odot} + M_{\rm r}$, 
 the extrinsic solar quadrupole moment ($J_2^{\rm ext}$), which can be measured gravitationally, becomes 
\begin{align}
    J_2^{\rm ext} =& (1 - \epsilon_{\rm r}) 
    J_2^{\rm int} + \epsilon_{\rm r} \left(\frac{\bar{R}}{R_{\odot}} \right)^2 J_2^{\rm ring} \label{eq:J2:ring_plus_sun}  \nonumber\\
    =& (1 - \epsilon_{\rm r}) J_2^{\rm int} + \epsilon_{\rm r} \left(\frac{1}{R_{\odot}^2} \right)\left[ \frac{1}{4}  \left(R_o^2 + R_i^2 \right) -\frac{h^2}{12} \right]\,, 
\end{align}
which can be solved for $\epsilon_{\rm r}$:
\begin{align}
    \epsilon_{\rm r} = \frac{2 \left( J_2^{\rm ext} - J_2^{\rm int} \right) }{\left( R/R_{\odot} \right)^2 - 2 J_2^{\rm int}} \,, \label{eq:ring:eps_r}
\end{align}
where
\begin{align}
    R \equiv& \sqrt{ \frac{R_o^2 + R_i^2}{2} -\frac{h^2}{6}} \,.\label{eq:def:ringR}
\end{align}
This reduces to $R = R_i = R_o$ for a very thin ring of  negligible height, \( h/R_i \approx 0 \). 
Here we assume $h/R_i \ll 1$ and note $R_i \ge R_{\odot}$. Defining $\delta J_2 \equiv J_2^{\rm ext}
- J_2^{\rm int}$ we rewrite Eq.~(\ref{eq:ring:eps_r}) as 
\begin{align}
    \frac{\delta J_2}{J_2^{\rm int}} 
    = &\, \epsilon_{\rm r} \left[\frac{1}{2 J_2^{\rm int}} \left(\frac{R}{R_{\odot}} \right)^2 - 1  \right] \approx \epsilon_{\rm r} \left[ \frac{1}{2 J_2^{\rm int}} \left(\frac{R}{R_{\odot}} \right)^2  \right] \,.
\end{align}
Recalling~\cite{Stenborg2018dustring} and thus the ring's proximity to Mercury's orbit 
with \( R_o \approx R_i \approx 0.38 \) AU 
we estimate: 
\begin{equation}
    \frac{\delta J_2}{J_2^{\rm int}} \approx \epsilon_{\rm r} \, (1.5 \times 10^{10}),
\end{equation}
implying that a $1\%$ difference 
in $J_2$, 
that is, 
$\delta J_2 \sim 10^{-2} J_2^{\rm int}$,
could be caused by $M_{\rm r} \sim 10^{-12}\, M_{\odot}$. Such a mass 
can contain a dark-matter component, though its value 
corresponds to that of a relatively large asteroid, 
significantly more massive than those 
more commonly found in the asteroid belt. We note, however, 
the dwarf planet Ceres, the most massive known 
object in the asteroid belt between Mars and Jupiter, has a mass of about 
$5\times 10^{-10} M_{\odot}$, whereas 
Jupiter's moon Amalthea has a mass of about 
$\left(1.04 \pm 0.08 \right)\times 10^{-12}\, M_{\odot}$~\cite{Anderson2005Amaltheas}. 
We now turn to concrete limits.

\begin{figure}[t]
    \centering
    \includegraphics[width=1\linewidth]{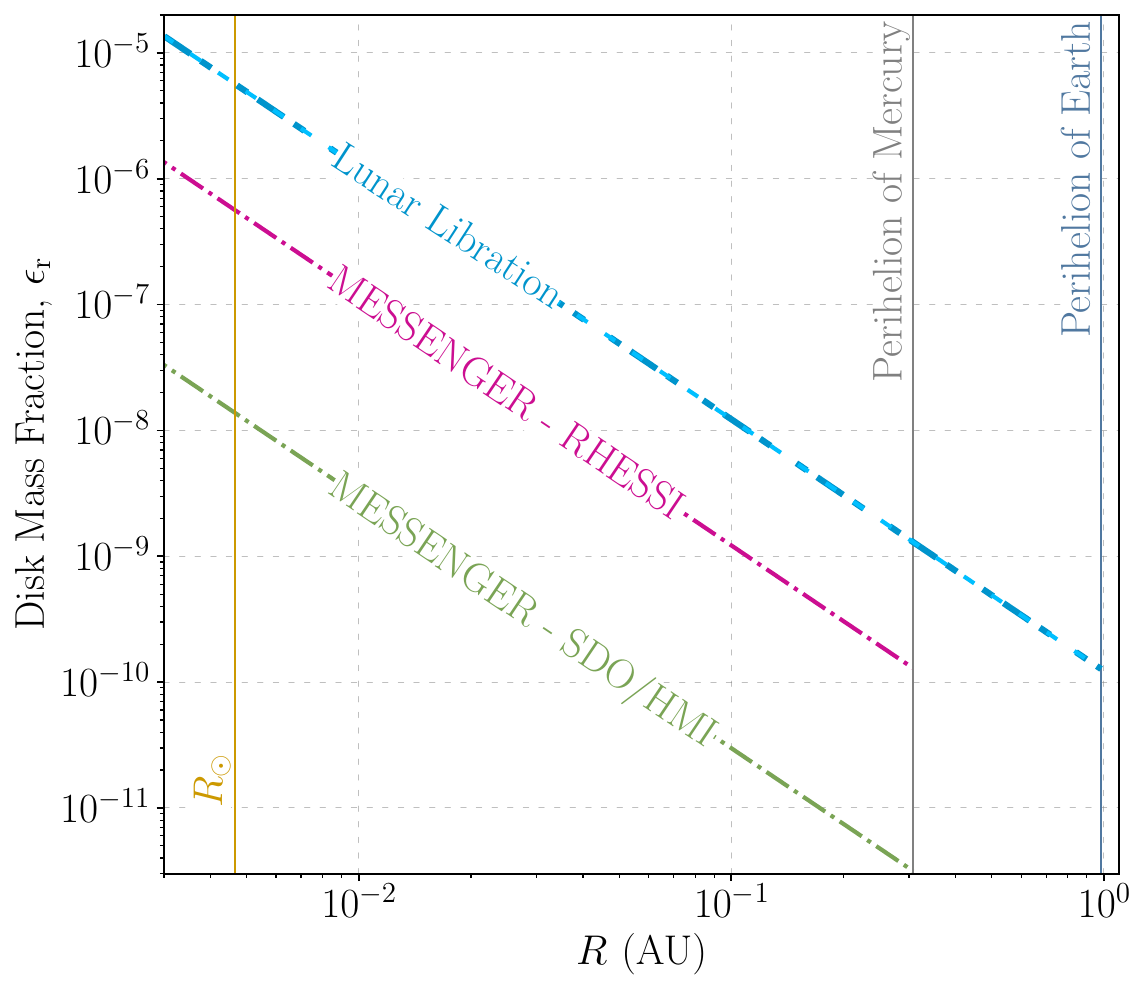}
    \caption{Limits at 95\% CL ($2\sigma$) 
    on the maximum mass fraction, $\epsilon_{\rm r} \equiv M_{\rm r} / M_{\rm tot}$, of a dark matter/dust ring or disk 
    around the Sun, derived from the difference in 
    gravitational and optical/helioseismological determinations of 
    the gravitational quadrupole moment $J_2$ as in 
    Eq.~(\ref{eq:epsilon_rsimple}), noting the maximum values of $\delta J_2$
    at $2\sigma$ in Eqs.~(\ref{eq:tinydeltaJ2}, \ref{eq:biggerdeltaJ2}), 
    plotted as dashed lines. 
The horizontal axis, $R$, is determined by 
the geometric dimensions of the dark matter/dust ring or disk, as defined in Eq.~\eqref{eq:def:ringR}. 
The 
dashed line corresponds to constraints inferred from the lunar libration limit on $J_2^{\rm ext}$~\cite{Bois1999QuadrupoleEarthMoon}.
    We refer to the text for all details.     }
    \label{fig:epsDM_vs_L:Ring}
\end{figure}

Given determinations of $J_2^{\rm ext}$ from orbital measurements $J_2^{\rm Orb}$ and of $J_2^{\rm int}$ from both optical and helioseismological measurements $J_2^{\rm Opt}$ and $J_2^{\rm Heli}$, 
we  constrain the mass in a 
circumsolar 
ring or disk through 
limits on 
$\epsilon_r$ in Eq.~\eqref{eq:ring:eps_r}.
Since we assign 
$M_{\odot}$ to mass
within $R_{\odot}$, 
practically 
$R/R_{\odot} > 1$, and 
since 
$J_2 \sim {\cal O}(10^{-7})$, we write 
\begin{equation}
\epsilon_r = 2 
\left( \frac{R_{\odot}}{R} \right)^2 \delta J_2
\,,
\label{eq:epsilon_rsimple}
\end{equation}
so that our ability
to determine 
$\delta J_2$ limits the 
maximum 
mass in the ring, regardless
of our confidence that 
$\delta J_2 >0$. 
We compute the 
maximum value of $\epsilon_r$ at 
95\% CL through 
different 
evaluations of the maximum value of 
$\delta J_2$ at $2\sigma$, 
and we display the 
results in 
Fig.~(\ref{fig:epsDM_vs_L:Ring}), noting that 
$R$ can range from within 
$R_{\odot}$ to the location of 
Mercury's orbit.

To accomplish this, 
we use $J_2^{\rm Orb}$ derived from the latest MESSENGER analysis~\cite{Genova2018MESSENGER}, given in Eq.~(\ref{eq:J2Genova_Orb}). This is then combined with either optical~\cite{Fivian2008SolarOblatMag}, as shown in Eq.~(\ref{eq:J2Opt}), or helioseismological measurements~\cite{Mecheri2021J2helio}, noted in Eqs.~(\ref{eq:J2_SDOHMI_CESAM}) and (\ref{eq:J2_SDOHMI_ASTEC}).
With the most 
significant difference, 
given 
in Eq.~(\ref{eq:J2diff}) (noting that if the enlarged error were used the allowed mass would be larger), we have 
\begin{equation}
    {\rm max} \left( \delta J_2 \right)\Big|_{2\sigma} = 
    0.069 \,, 
    \label{eq:tinydeltaJ2}
\end{equation}
whereas combining Eq.~(\ref{eq:J2Genova_Orb}), 
using either the $\beta=1\,;\, \eta=0$ constrained (GR) or 
unconstrained value, 
with the optical result of 
Eq.~(\ref{eq:J2Opt}) yields 
\begin{equation}
    {\rm max} \left( \delta J_2 \right)\Big|_{2\sigma} = 2.82\, (2.83 )
     \,, 
     \label{eq:biggerdeltaJ2}
\end{equation}
respectively. 
We report the limits on $\epsilon_r$ with $R$ at 95\% CL ($2\sigma$) that come out, using the 
GR-constrained orbital result  in 
Fig.~\ref{fig:epsDM_vs_L:Ring}. 
Furthermore, 
we also depict the 
constraint on $\epsilon_r$ that follows 
from the limit $J_2^{\rm ext} < 3 \times 10^{-6}$~\cite{Bois1999QuadrupoleEarthMoon}, 
determined by demanding that the 
lunar librations, which are modelled and observed from the analysis of 
lunar laser ranging (LLR) measurements of the Earth-Moon 
distance~\cite{Dickey1994LLR}, do not exceed $3\sigma/2$ of the LLR residuals.  
We note that this limit is sufficiently weak 
that the impact of the use of 
$J_2^{\rm Opt}$ in Eq.~(\ref{eq:J2Opt}) or $J_2^{\rm Heli}$ from
combining Eqs.~(\ref{eq:J2_SDOHMI_CESAM},\ref{eq:J2_SDOHMI_ASTEC})
is indistinguishably small. 

Thus far we have limited $\epsilon_r$ supposing that a dark disk or
ring exists. Now, we assess the likelihood of this scenario using Eq.~(\ref{eq:darkdisk}) and the observed values of $J_2^{\rm ext}$ and $J_2^{\rm int}$. Adopting a Gaussian prior for $\delta J_2$, the relevant posterior probability is given by~\cite{james2006statistical}
\begin{equation}
    P(\delta J_2 > 0 \,|\, J_2^{\rm ext}, J_2^{\rm int}) = 1 - F_{\delta J_2}(z), \label{eq:ring:posterior}
\end{equation}
where \( F_{\delta J_2}(z) \) is the cumulative distribution function (CDF) of the standard normal distribution, 
\begin{equation}
    F_{\delta J_2} (z)
    = \frac{1}{\sqrt{2\pi}}
    \int_z^\infty dt \exp(-t^2/2)\,, 
\end{equation}
and $z = -\mu_{\rm post} / \sigma_{\rm post}$. Here, $\mu_{\rm post}$ and  $\sigma_{\rm post}$ are the posterior mean and standard deviation, respectively. For a non-informative (flat) prior where the standard deviation of the prior ($\sigma_{\rm prior}$) tends towards infinity, $\mu_{\rm post}$ and $\sigma_{\rm post}$ simplify to yield
\begin{equation}
z \approx \frac{\mu_{J_2^{\rm int}} - \mu_{J_2^{\rm ext}}}{\sigma_{J_2^{\rm ext}} + \sigma_{J_2^{\rm int}}}.
\end{equation}
 The deviation of the computed posterior probabilities, as formulated in Eq.~(\ref{eq:ring:posterior}), from one are presented for six distinct cases in Table~\ref{tab:ring:posterior}. We observe that even in the unconstrained cases that the probability that a disk  exists is high,
although our concern regarding solar
model uncertainties remains. 
\begin{table}[h!]
\centering
\begin{tabular}{|c|c|c|}
\hline
\diagbox{int}{ext} & constrained & unconstrained \\ 
\hline
Opt & 0.210 & 0.221 \\
\hline
Heli & $1.0\times10^{-21}$ & 0.091 \\
\hline
Heli-All & $2.4\times 10^{-17}$ & 0.089 \\
\hline
\end{tabular}
\caption{\label{tab:ring:posterior}The deviation of the disk structure's presence likelihood from one, $1-P(\delta J_2>0)$, evaluated using Eq.~(\ref{eq:ring:posterior}). We compared the GR-constrained (unconstrained) $J_2^{\rm ext}$ values from MESSENGER data, as given by Eq.~(\ref{eq:J2Genova_Orb}), against the best optical and helioseismological determinations of $J_2^{\rm int}$, given by Eqs.~(\ref{eq:J2Opt}, \ref{eq:J2_finalHelio},  
\ref{eq:J2heli-all}), 
respectively. We refer to the text for all details.}
\end{table}

{\it Limiting the mass of a non-luminous spherical halo:}

For a spherical halo, setting $J_2^{\rm ring} = 0$ in Eq.~(\ref{eq:J2:ring_plus_sun}) yields
\begin{align}
    \epsilon_{\rm s} = \frac{J_2^{\rm int} - J_2^{\rm ext}}{J_2^{\rm int}} \,, \label{eq:sphere_plus_sun:eps_s}
\end{align}
in which we denote the fractional mass of the spherical halo by \( \epsilon_{\rm s} \equiv M_{\rm s} / M_{\rm tot} \). We apply the Feldman-Cousins~\cite{Feldman:1997qc} method to establish limits on $\epsilon_{\rm s}$ given pairs of $(J_2^{\rm ext}, J_2^{\rm int})$ values. The resulting limit on $\epsilon_{\rm s}$, derived by comparing the constrained (unconstrained) MESSENGER $J_2^{\rm ext}$ values in Eq.~(\ref{eq:J2Genova_Orb}) with $J_2^{\rm Heli}$
from Eq.~(\ref{eq:J2_finalHelio}), is given by
\begin{equation}
   \epsilon_{\rm s}\Big|_{2\sigma} = 2.5 \times 10^{-4} \, (8.3 \times 10^{-3}). \label{eq:sphere:limit}
\end{equation}

\begin{figure}[t]
    \centering
    \includegraphics[width=1\linewidth]{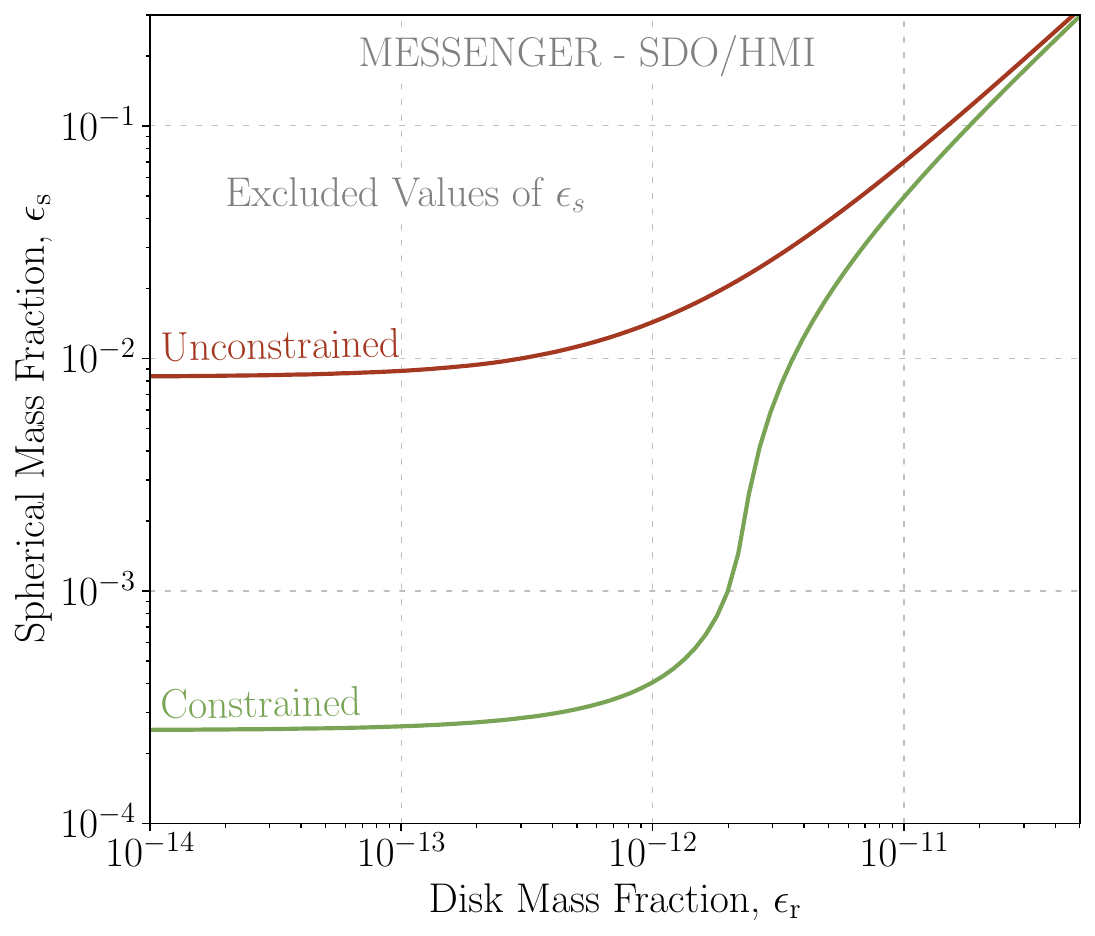}
    \caption{Limits at 95\% CL ($2\sigma$) on the mass fraction of a spherical dark matter halo surrounding the Sun, denoted as $\epsilon_{\rm s} \equiv M_{\rm s} / M_{\rm tot}$, plotted against the mass fraction, $\epsilon_{\rm r} \equiv M_{\rm r} / M_{\rm tot}$, of a dust ring orbiting within Mercury's path. We input the values from 
    Eqs.~(\ref{eq:J2Genova_Orb}, \ref{eq:J2_finalHelio}) into Eq.~(\ref{eq:sphere_plus_ring_plus_sun:eps_s}). The GR-constrained and unconstrained cases are shown in green and red, respectively. We refer to the text for all details. 
    }
    \label{fig:eps_s_vs_eps_r}
\end{figure}

We now take the existence of Mercury's circumsolar dust ring into account~\cite{Pokorny2023Mercuryimpact}. The portion of the dust ring that falls within the orbit of Mercury would increase $J_2^{\rm ext}$, such that Eq.~(\ref{eq:sphere_plus_sun:eps_s}) changes to
\begin{align}
    \epsilon_{\rm s} = \frac{(1 - \epsilon_r) J_2^{\rm int} - J_2^{\rm ext} + (\epsilon_r/2) \left(R / R_{\odot}\right)^2}{J_2^{\rm int}} \,, \label{eq:sphere_plus_ring_plus_sun:eps_s}
\end{align}
which would weaken the bounds on $\epsilon_s$ in Eq.~(\ref{eq:sphere:limit}) as shown in Fig.~\ref{fig:eps_s_vs_eps_r}, where we set the dust ring's radius to $R\approx 0.31$  AU.

\
\section{Implications for dark-matter models}
\label{sec:dm_impl}

We have found that existing determinations of $J_2^{\rm Heli, Opt}$ ($J_2^{\rm int}$) 
and $J_2^{\rm Orb}$ ($J_2^{\rm ext}$) from the MESSENGER mission 
favor the existence of a 
non-luminous disk within Mercury's orbit, with Table \ref{tab:ring:posterior}
providing the likelihood that it does {\it not} exist. Although 
the significance of our claim depends on the results we
pick, the limits we have found on its mass, as shown in Fig.~\ref{fig:epsDM_vs_L:Ring}, grossly exceed 
the contributions we would expect from 
the cosmological dark matter density and/or the dust ring that has
been discovered near 
the path of Mercury's orbit~\cite{Stenborg2018dustring}.  
The observed dust ring, assessed at an overdensity of
$\approx 3-5
\%$~\cite{Stenborg2018dustring}, 
is estimated to be 
located between
heliocentric distances (radii) of $0.356271\, \rm AU$ and $0.400308 \, \rm AU$, 
with a mass of 
$(1.02 - 4.05) \times 10^{12\pm (\approx 1)} \,\rm kg$~\cite{Pokorny2023Mercuryimpact}. 
Dust, in grains 
ranging in size from some 10 (100) $\mu\rm m$ 
to 1 cm in diameter with a density of $2000\, \rm kg  / m^{3}$,  
is generally an expected component~\cite{Pokorny2023Mercuryimpact}. 
Dust within our Zodiacal Cloud is observed to be depleted very close 
to the Sun~\cite{Stenborg2021nodustnearsun}, 
and we follow
\cite{Pokorny2023Mercuryimpact} in assuming this effect can be neglected
beyond a radius of $0.05\, \rm AU$
to estimate that the 
mass in dust within 
Mercury's orbit 
to be 
 no more than about 
$7 \times 10^{15 \pm (\approx 1)} \,\rm kg$, or 
$3\times 10^{-15 \pm (\approx 1)} \, \rm M_\odot$. This is still 
a few orders
of magnitude smaller than our most stringent 
bound on the maximum mass within the non-luminous disk, though we can expect
other massive objects to appear as well. 
For example, a spherical object of the same 
composition as dust with a radius of 2.5 km would have a mass of 
$1.3\times 10^{14}\rm kg$, so that some 100 of them would be needed to 
contribute a mass of $10^{16}\rm kg$, crudely comparable to that in dust.
We note the solar system
survey from {\it Gaia} Data Release 3
reveals 432 near-Earth objects with a diameter of 5 km or less~\cite{Tanga2023solarsystemGaiaDR3}, though the total 
number and mass of such objects within Mercury's orbit is unknown. 
Much lighter objects, such as micrometeroids and others,  
may also appear in an appreciable way~\cite{Sommer2023micrometeroid,Taylor2024darkcomet}. 
Thus there could well be missing matter, of differing origins, 
within Mercury's orbit --- and we now turn to how 
some measure of it could be {\it dark matter}. 
 
The cosmic dark-matter density 
contributes only $10^{-19}\rm M_{\odot}$ of the mass within Mercury's
orbit, and 
local effects could 
modify that total mass. 
For example, a macroscopic dark object, such as a PBH, 
could exist. Interestingly, 
in the mass range of  $(10^{-15} - 10^{-11}) \, \rm M_\odot$, 
PBHs could constitute 
all of the dark 
matter~\cite{Green2024PBHdarkmatterrev}, though X-ray studies can 
offer further constraints~\cite{Tamta2024PBHXray}. Nevertheless, 
a PBH 
in the expected mass range 
could be in transit through (or contained within)
the solar system within
Mercury's orbit, where we note \cite{Tran2023PBHdarksolarsystem} for solar-system 
transit rate estimates.

Ultra-light dark matter could also play a role. 
To that end, we consider the generic halo formation mechanism
of \cite{Budker2023gravatom}, arising from dark-matter self-interactions
in the gravitational field of a massive celestial body, 
such as the Sun. We note that ultra-light dark matter models with a particle mass of $10^{-23} \, {\rm eV} < m <10^{-17}\, {\rm eV}$ 
can produce dense
solitonic cores in the Galactic Center,
in which case the 
particle mass is constrained by the 
observation of the motion of the star S2~\cite{DellaMonica2023ULDMboundSgA}. 
However, these constraints do not apply to the models considered here. The gravi-atom picture is promising in that a strong 
enhancement 
of the dark-matter density is possible in particular
regions of parameter space; for example, this model can 
yield a spherical halo about the Sun with an overdensity of 
$\delta\rho_{\rm dm} \equiv \rho_{\rm crit}(0) / \rho_{\rm dm} \simeq 7\times 10^3$ for an
axion dark-matter candidate with a mass of $10^{-14}\, \rm eV$ and 
an axion decay constant of $5\times 10^7 \, \rm GeV$, with the
self-interaction strength $\lambda$ given by $\lambda = - m^2 / f_a^2$. 
The estimated extra mass this halo contributes within Mercury's mean orbit is more 
than fifty times smaller~\cite{Budker2023gravatom} than the 
``extra mass'' ephemerides constraint of 
$\rho_{\rm dm} < 9.3 \times 10^{-18} \,\rm g/cm^3$~\cite{Pitjev2013constraints}, which yields an enclosed mass of $< 7\times 10^{12} \,\rm kg$ if $\rho_{\rm dm}$
were uniform. The constraint of \cite{Pitjev2013constraints} includes a solar 
$J_{2}$ of a suitable size and error~\cite{Pitjeva2014Ephemerides}, so that this mass limit is not at odds with our own analysis. Nevertheless, 
the spherical symmetry of this dark halo 
is at odds with the 
evidence we have found for an extended disk-like object. 
However, we can adapt the gravi-atom 
mechanism~\cite{Budker2023gravatom} to our case by noting that the 
central mass $M$ need only be much more massive than that of the dark-matter candidate $m$, 
with a halo of radius $R_\ast$ that exceeds that of the central mass. 
Particularly, we evaluate the possibility that halos form around clumps of 
conventional matter in the disk, possibly yielding
a number of $\ll 1 \rm AU$-sized halo states. 
The total dark mass in the disk would then be given by the sum of the masses 
associated with each of the halo bound states. 

Translating the gravi-atom estimates in $M$ and $m$ 
to kg and eV scales 
is possible because the size of the bound
state, as well as the possible overdensity, are each determined by $M m^2$, 
where $M_\odot (10^{-14} \rm eV)^2 \approx 200\, kg\, eV^2$.  
Here, successful halo formation in the gravi-atom picture requires a
suitably sized $R_*$, a wave-like dark-matter candidate so that 
the de Broglie wavelength $\lambda_{\rm dB} $ exceeds the dark matter
interparticle spacing, and a maximum halo density $\rho_{\rm crit}$ (at its
center) that significantly exceeds $\rho_{\rm dm}$ ($\delta \rho_{\rm dm} \gg 1$). There are also dynamical 
considerations, in that dark-matter capture should exceed stripping, 
making $\xi_{\rm foc} \equiv \lambda_{\rm db} / R_{\ast} \gtrsim 1 $, 
with the density reaching 
$\rho_{\rm crit}$ 
within about 5 Gyr, the lifetime of the Solar system~\cite{Budker2023gravatom}. 
There are also independent constraints on the axion from direct 
searches if it couples to photons or nucleons, and a favored parameter
space of $m \in (0.7 - 700)\, \mu \rm eV$ with $f_a \in (2\times 10^{13} 
- 2\times 10^{10})\,\rm GeV$, if it ought also solve the strong CP problem of QCD~\cite{OHare2024reviewaxioncosmo}. 
Thus the solar halo suggested 
in \cite{Budker2023gravatom} is unlikely to be generated by a QCD axion, nor
can that axion couple to photons. 
These features will 
continue to 
bear out in the examples 
to which we now turn. 

Different sorts of constituents could potentially act as the nucleus of a small 
dark-matter halo. Dust, 
for example, is thought to exist in grains  
ranging in size from some 10 (100) $\mu\rm m$ 
to 1 cm in diameter with a density of $2000\, \rm kg  / 
m^{3}$~\cite{Pokorny2023Mercuryimpact}, and larger clumps of matter should also occur.
Here, we explore different
scenarios for dark-halo formation, considering 
central masses comprising
of (i) a small asteroid with $M=1.3\times 10^{14} \,\rm kg$
(a $2.5\, \rm km$-radius rock of 
density $2000\, \rm kg/m^3$), of (ii) a much-denser asteroid with $M=10^{16} \,\rm kg$, 
or of (iii) Mercury itself
with $M=3.3\times 10^{23}\, \rm kg$, with a radius of 
$2440 \,\rm km$, so that 
$R_\ast$ must be larger
than the noted radius in 
each case. In these cases $M \ll M_{\odot}$, and we might 
suppose an initial encounter of a dark-matter particle with the Sun 
produces  a population of dark-matter particles that satisfy 
$v\ll v_{\rm halo} \approx 240 \, \rm km/s$. We do not think it is possible
to generate gravi-atoms with dust grain cores so that $\xi_{\rm foc} \gtrsim 1$
is satisfied.

As detailed in 
Appendix~\ref{sec:app:graviatom}, 
the chosen scenarios for $M$ can satisfy the 
noted constraints under the 
following conditions: 
(i) $M = 1.3 \times 10^{14} \, {\rm kg} ; m = 1.0\times 10^{-2}\, {\rm eV} ; 
    f_a = 1.0\times 10^{1} \, {\rm GeV}$ for which 
    the radius of the ground state is 
    $R_\ast = 4.0  \, {\rm km}$, the gravitational
    fine-structure constant
    $\alpha=4.9\times 10^{-9}$, 
    and the dark-matter 
    self-coupling is 
    $\lambda = 1.0\times 10^{-24}$ --- and the estimated maximum overdensity is
    $\delta \rho_{\rm dm} = 1.2 \times 10^6$. If this can be attained, 
    then $M_{\rm enc, max} 
    \approx \delta\rho_{\rm dm} \rho_{\rm dm} 4\pi R_{\ast}^3 /3 = 2.4 \times 10^{-4} \, \rm kg$. If
        $v_{\rm halo}$ characterizes the dark-matter particle speed, then 
   \begin{eqnarray}
     \xi_{\rm foc} = 3.8\times 10^{-5} ; \tau_{\rm rel} = 9.1\times 10^8\, {\rm Gyr}
\end{eqnarray}
but with $v\approx 8 \,\rm m/s$ (just for illustration), $\xi_{\rm foc} \rightarrow 1.1$ and 
$\tau_{\rm rel} \rightarrow 1.0\, \rm Gyr$. However, we do not think there would be 
enough rocks in situ for this solution, if attainable, to generate a significant
contribution to the disk mass. 

We compare this to a heavier object that is six-times denser, finding 
(ii) $M = 1.0 \times 10^{16} \, {\rm kg} ; m = 6.0 \times 10^{-5} \, {\rm eV} ; 
    f_a = 1.0 \times 10^{1} \, {\rm GeV}$ for which 
    the radius of the ground state is 
    $R_\ast = 1.5 \times 10^{3}  \, {\rm km}$, the gravitational
    fine-structure constant
    $\alpha=2.3\times 10^{-9}$, 
    and the dark-matter 
    self-coupling 
    $\lambda = 3.6\times 10^{-29}$. Here 
    $\delta \rho_{\rm dm} = 9.6$ and
$M_{\rm enc, max} \approx 8.8 \times 10^{-2} \, \rm kg$ with 
    \begin{eqnarray}
     \xi_{\rm foc} = 1.8\times 10^{-5} ; \tau_{\rm rel} = 2.0 \times 10^{2} \, {\rm Gyr}
\end{eqnarray}
using $v_{\rm halo}$. By making the central mass denser, we see that 
more mass can be stored in the halo, but here, too, it would not seem possible
to generate a significant contribution to the disk mass. 

Finally we consider the possibility of a halo about Mercury itself, and 
we consider 
    (iii) 
      $ M = 3.3 \times 10^{23} \, {\rm kg} ; m = 9\times 10^{-11} \, {\rm eV} ; 
    f_a = 1.0\times 10^{7} \, {\rm GeV}$ so that 
    $R_\ast = 0.13 \, {\rm AU}$,     $\alpha=1.1\times 10^{-7}$, 
    $\lambda = 8.1\times 10^{-53}$. Here 
    $\delta \rho_{\rm dm} = 5.3 \times 10^4$ and
$M_{\rm enc, max} \approx 1.2 \times 10^{15} \, \rm kg$ with 
    \begin{eqnarray}
     \xi_{\rm foc} = 8.7\times 10^{-4} ; \tau_{\rm rel} = 6.6 \times 10^{8} \, {\rm Gyr}
\end{eqnarray}
for $v_{\rm halo}$, and
if $v\approx 180\, \rm m/s$, 
 $\xi_{\rm foc} \rightarrow 1.2$ and 
$\tau_{\rm rel} \rightarrow 3.7\times 10^2\, \rm Gyr$ 
--- and 
the dynamical constraints can be reasonably well satisfied. 
In this case a contribution to the mass of the disk
could be as large as $6\times 10^{14} \, \rm kg$, though a deviation from sphericity, presumably from gravitational focussing, would be needed to impact the perihelion precession. Nevertheless, it appears 
possible to generate macroscopic
contributions to a non-luminous disk from ultra-light dark matter, even if 
we have not yet found 
an example through 
which it can grossly 
dominate the expected non-luminous mass.

\section{Summary} 
\label{sec:future}

In this paper 
we have considered different 
determinations 
of the Sun's gravitational 
quadrupole moment $J_2$, 
carefully comparing optical assessments of the Sun's shape with interferences  from orbital observations, particularly using MESSENGER studies of Mercury. 
The pattern of observations, 
considered broadly, 
favor the pattern 
\begin{equation}
J_2^{\rm Orb} > J_2^{\rm Opt} \,,\, J_2^{\rm Heli} \, 
\end{equation} 
given in 
Eq.~(\ref{eq:darkdisk}), 
which speaks to 
the existence
of a non-luminous disk. 
Particularly, if we compare
the orbital results assuming 
GR and 
using MESSENGER data~\cite{Genova2018MESSENGER}, 
Eq.~(\ref{eq:J2Genova_Orb}), 
with contemporaneous 
helioseismological 
results from a space-based
observatory~\cite{Mecheri2021J2helio}, 
Eq.~(\ref{eq:J2_finalHelio}), we find 
the difference, given in Eq.~(\ref{eq:J2diff}), 
\begin{equation}
J_2^{\rm Orb}|_{\beta=1; \eta=0} - J_2^{\rm Heli}
= 0.057 \pm 0.006  \,\, (\pm 0.020)\,
\end{equation}
is greater than zero 
with a significance of $\simeq 3\sigma$.
We have also explored how alternative helioseismological error estimates and outcomes impact this conclusion, and we find
support for a 
significantly positive result.
Consequently,  
we have developed constraints 
on the total 
mass and mass distribution of that 
non-luminous disk, 
and we claim that the $2\sigma$-limit on its
maximum mass 
is no less than some $10^{-12}\rm M_{\odot}$, where increasing
the error in the 
$J_2$ difference gives rise to a larger maximum-mass limit. 
We have also 
carefully considered its 
possible components, and 
we found that the missing
(non-luminous) mass is sufficiently large that 
conventional sources of mass, 
from dust, from asteroids and 
meteoroids of all sizes, would
not seem able to explain it. 
We also believe the total mass of these various components to be poorly determined. 
We nevertheless suppose that dark matter could also contribute to its mass. Thus
our study points to a 
possible ``missing mass'' problem within Mercury's orbit. 
A macroscopic
dark matter candidate, 
such as a PBH of a mass 
typical of an asteroid, 
is one concrete possibility. 
We have also considered how a particular 
ultra-light dark matter model with self-interactions~\cite{Budker2023gravatom}
can generate 
a massive dark 
halo about Mercury itself, 
potentially
contributing to the 
appearance of a massive ring in the path 
of Mercury's orbit. 

We anticipate that the BepiColombo  
mission to Mercury, to arrive in 2025, can
refine the results
from the earlier MESSENGER mission, that we
have exploited in this paper, with further 
precision measurements of Mercury's orbit
and magnetic properties offering the 
prospect of confirming and refining our 
findings --- or not. 
Particularly, precise assessments of short time-scale perturbations on Mercury's orbit should constrain massive exotic objects in transit through the solar system, such as the PBHs candidates discussed in \cite{Tran2023PBHdarksolarsystem}.
Although the dark-matter
particles we have particularly considered
would not seem to be axions, we expect
that magnetic field studies may well be 
discriminating nonetheless, 
as they can probe the possibility of macroscopic dark-matter objects with a slight electric charge. 
Additional studies of dust are also 
planned~\cite{Kobayashi2020dustprobefuture}. 
Ultimately, too, we expect these studies to 
have implications for the precision of future 
relativistic general relativity tests~\cite{Will2018newGRperihelion}. 

To confirm and refine our findings we emphasize that the following probes 
are possible in the near future:

\begin{itemize}
    \item Detected perturbations in Mercury's orbit can speak
to the existence of ultraheavy dark matter. 
    \item Studies of light reddening in the inner solar system can be
    used to separate a dusty, non-luminous component from dark matter. 
    \item The JUNO neutrino experiment is poised to measure CNO neutrinos~\cite{JUNO2024CNO} with higher precision than BOREXINO and 
can test the latter's inference of a nonhomogeneous zero-age
Sun, which supports the existence of an early
protoplanetary disk, some of which may still remain. 
\end{itemize}

Thus we are hopeful that the possible
``missing mass'' problem we have 
uncovered can be clarified 
in the relatively near future.

\begin{acknowledgments}
We thank Lance Dixon, Antonio Genova, Roni Harnik, Dan Hooper, Russell Howard, Mrunal Korwar, and Guillermo Stenborg  for helpful correspondence and comments, 
and we are grateful to the anonymous referee for bringing the work of \cite{Konopliv2020Icar..33513386K} to our attention. 
S.G. would like to thank the Theoretical Division at Fermilab for lively 
hospitality and to acknowledge the Universities Research Association for support during
a sabbatical visit in which this project was initiated. 
G.F.S.A. would like to thank the hospitality of the Fermilab Theory Group. G.F.S.A., P.M., and M.Z. are grateful to the Center for Theoretical Underground Physics and Related Areas (CETUP*), The Institute for Underground Science at Sanford Underground Research Facility (SURF), and the South Dakota Science and Technology Authority for their hospitality and financial support. 
That stimulating environment was invaluable to us. 

S.G. and M.Z. also acknowledge support from the U.S. Department of Energy, Office of Science, Office
of Nuclear Physics 
under contract DE-FG02-96ER40989. 
P.M. is supported by Fermi Research Alliance, LLC under Contract No. DE-AC02-07CH11359 with the U.S. Department of Energy, Office of Science, Office of High Energy Physics.
G.F.S.A. is fully financially supported by Fundação de Amparo à Pesquisa do Estado de São Paulo (FAPESP) under Contracts No. 2022/10894-8 and No. 2020/08096-0. 
\end{acknowledgments}

\appendix
\section{Oblateness and Its Gravitational Effects}
\label{sec:app:oblateness}

Here we show how a relationship between the solar oblateness and its gravitational moments can be established under the assumption that the solar surface constitutes an equipotential surface~\cite{Zeipel1924EqRotSystems, Dicke85Measurement83}, i.e., $\Phi(R_e) = \Phi(R_p)$. Assuming further that the Sun's rotation ($\Omega$) depends solely on the distance from its rotation axis, $l = r \sin\theta$, and disregarding magnetic stresses, the effective potential can be expressed as~\cite{Dicke1970oblatenessJ2ApJ}
\begin{equation}
\Phi \equiv \phi_{i} - \phi_{\Omega} = \phi_{i} - \int_{0}^{l_{\odot}} l\, \Omega^2(l)\, dl, \qquad l = r\sin\theta, \label{eq:effpot}
\end{equation}
where $\phi_{i}$ and $\phi_{\Omega}$ represent the internal potential of the Sun and the effective potential due to rotation, respectively. The continuity of the potential across the Sun's surface yields $ \phi_{i}^{\rm surf} = \phi_{o}^{\rm surf}$, where $\phi_{o}$ is the gravitational potential outside the Sun, given by Eq.~\eqref{eq:gravphio}. 
The equipotential condition $\Phi(R_e) = \Phi(R_p)$ results in
\begin{equation}
    \frac{\Delta_{\odot}}{\eta_p^2 \left( 1 + \frac{\Delta_{\odot}}{\eta_p}\right)} + \sum_n\, \frac{J_n}{\eta_p^{n+1}} \left( \frac{P_n(0)}{\left( 1 + \frac{\Delta_{\odot}}{\eta_p}\right)^{n+1}} - 1 \right) + \frac{R_{\odot}\, \phi_{\Omega}(R_e)}{G\, M_{\odot}} = 0, \label{eq:equipot}
\end{equation}
where $\eta_{e, p} \equiv R_{e, p} / R_{\odot}$, and $\eta_e$ replaced by $\eta_e = \Delta_{\odot} + \eta_p$. Noting that $(\eta_e + \eta_p)/2 = 1 + \mathcal{O} (\Delta_{\odot}^2)$, $\eta_p$ expanded to linear order in $\Delta_{\odot}$ is given by $\eta_p \approx 1 - \Delta_{\odot}/2$. Hence, Eq.~\eqref{eq:equipot} expanded to linear order in $\Delta_{\odot}$ and up to the octopole moment ($n=4$) is
\begin{equation}
   \Delta_{\odot}^{(4)} = \frac{\left(16 R_{\odot}\, \phi_{\Omega}(R_e) / G M_{\odot}\right) - 2 \left(8 J_1+12 J_2+8
   J_3+5 J_4\right)}{4 \left(4
   J_1+3 J_2+8 J_3-4\right)+55
   J_4}. \label{eq:equipot:linear}
\end{equation}
Given that all the multipoles are expected to be smaller than the monopole term, $J_n \ll 1$, the oblateness in Eq.~\eqref{eq:equipot:linear} can be approximated by keeping only the linear terms:
\begin{align}
   \Delta_{\odot}^{(4)} \approx J_1 + \frac{3 J_2}{2} + J_3 + \frac{5 J_4}{8} - \frac{R_{\odot}\, \phi_{\Omega}(R_e) }{G M_{\odot}}, \label{eq:equipot:linear:smallJ}
\end{align}
which in the case of a rigidly rotating body reduces to Eq.~\eqref{eq:DickeJ2plus}.

\section{Gravitational Quadrupole Moments}
\label{sec:app:tables}

We have performed an extensive review of the previous determinations of the solar gravitational 
quadrupole moment. 
There are three distinct 
ways by which the Sun's shape can be inferred: 
through the optical measurement of its oblateness
(Table \ref{tab:J2:opt}), 
through helioseismology (Table \ref{tab:J2:helio}),  and through measurements of planetary orbits (Table~\ref{tab:J2:orb}
and Table~\ref{tab:J2_Orb_extended}). We base our analysis on a few values from these Tables,  and we detail our motivations in doing
so here. First, among the optical measurements reported in Table~\ref{tab:J2:opt}, only results \#14--\#17 
take into account corrections for surface magnetism, and 
\cite{Fivian2008SolarOblatMag} (\#14) provides the most precise determination 
to date, though we
note \cite{Kuhn2012precisesolshape} (\#17) for further discussion of that
earlier work.

The helioseismological measurements 
in
Table~\ref{tab:J2:helio} 
have been categorized based on the instruments used. 
The values reported in 
\cite{Mecheri2021J2helio}, 
namely,  
\#10--\#11 and \#17--\#18
in Table \ref{tab:J2:helio} 
figure prominently in our 
analysis because of their 
improved analysis framework 
and the extensive helioseismological data
sets they employ. 
We can also consider the results for each distinct instrument and determine the average value and standard deviation for each case, to find using Table \ref{tab:J2:helio} that 
\begin{align}
  J_2^{\rm SoHO/MDI}  =& 2.135 \pm 0.203\,, \quad 
  \\
  J_2^{\rm GONG}  =& 2.167 \pm 0.019\,, \quad 
  \\
  J_2^{\rm SDO/HMI}  =& 2.214 \pm 0.002\,. \quad 
\end{align}
We use these to 
calculate a combined average value of $J_2$ of these
independent measurements, 
yielding 
\begin{equation}
   \langle     J_2^{\rm Heli} \rangle \Big|_{\rm all}  = 2.213 \pm 0.002 \,.
\end{equation}
Thus we see that we recover a result compatible with Eq.~(\ref{eq:J2_finalHelio}) through
completely different means, supporting our earlier analysis, 
albeit the HMI results have the smallest dispersion.

Lastly, for orbital determinations we employ only
the most recent 
$J_2$ analysis
using MESSENGER data~\cite{Genova2018MESSENGER}. 
This work does take into account the ELT correction and makes a simultaneous fit of PPN parameters in their $J_2$ determination. These are important additions, and we have to make sure we are comparing $J_2$ assessments that take into account the same effects. Table.~\ref{tab:J2:orb} serves this purpose as we opt to report only those values that consider these additions in their $J_2$ determination. For completeness, in Table~\ref{tab:J2_Orb_extended} we compile several orbital assessments for comparison indicating whether they included either the ELT effect, a simultaneous fit of PPN parameters, or both.

\setlength{\tabcolsep}{15pt}
\renewcommand{\arraystretch}{1.3}

\begin{widetext}

\begin{table*}[!htbp]
    \centering
    \begin{threeparttable}
    \label{tab:summary_J2_Optical}
     \begin{tabularx}{0.63\textwidth}{@{} l c S[table-format=-3.2] S[table-format=-3.2]
     c @{}}
    \toprule
    \multicolumn{5}{c}{Optical $J_2$ measurements} \\
    \toprule
    \# & {$\Delta_{\odot}(\times 10^{-6})$} & {$J_2 (\times 10^{-7})$} & {$\pm J_2 (\times 10^{-7})$} & Reference \\
    \midrule
    1 & $50 \pm 7$ &  279 & 47 & Dicke, 66~\cite{Dicke1967SolarOblat}\\
    2 &$9.6 \pm 6.6$ & 9.91 & 43.8& Hill, 73~\cite{Hill1975VSO}\\
    3 &$20.0 \pm 1.5$ & 79.4 & 9.73& Dicke, 83~\cite{Dicke85Measurement83}\\
    4 &$5.8\pm 1.4$ & -15.1 & 9.03 &Dicke, 84~\cite{DickeKuhnLibbrecht1986OblatMeasurement}\\
    5 &$15.2 \pm 2.3$ & 47.4 & 15.3 & Dicke, 85~\cite{Dicke1987J2Variable}\\
    6 &$ 14.4\pm 1.4$ & 41.9 & 9.03 & Beardsley, 83~\cite{Beardsley1987VSO}\\
    7 &$5.6 \pm 6.3$ & -16.7 & 42.0 & Maier, 90~\cite{Maier1992VSO} \\
    8 &$4.3 \pm 2.0$ & -25.3 & 13.3 & Egidi, 92~\cite{Egidi2006VSO}\\ 
    9 &$8.5 \pm 2.1$ & 2.67 & 14.0 & Egidi, 93~\cite{Egidi2006VSO}\\ 
    10 &$8.6 \pm 1.4$  & 3.33 & 9.33 & Egidi, 94~\cite{Egidi2006VSO} \\
    11 &$10.3 \pm 1.9$ & 14.7 & 12.7 &Egidi, 95~\cite{Egidi2006VSO} \\
    12 & $9.0 \pm 2.9$ & 6.44 & 19.5 & Emilio, 97~\cite{Emilio2007ChangingSolarShape}\\
    13 & $19.7 \pm 1.9$ & 77.3 & 13.2 & Emilio, 01~\cite{Emilio2007ChangingSolarShape}\\
    14 &$8.35 \pm 0.15$ & 1.46 
    & 1.0
    & Fivian, 02--08~\cite{Fivian2008SolarOblatMag}\\
    15 & $8.8 \pm 0.3$ & 4.3 & 2.1 & Irbah, 11~\cite{Irbah2014PICARD}\\
    16 & $8.19 \pm 0.33$ & 0.55 & 2.22 & Meftah, 10--11~\cite{Meftah2015PICARDoblat}\\
    17 & $7.5 \pm 0.5$ & -4.0  & 3.4 & Kuhn, 10--12~\cite{Kuhn2012precisesolshape}\\ 
    \bottomrule
    \end{tabularx}
    \caption{\label{tab:J2:opt} Summary of optical $J_2$ measurements. To translate from visual oblateness $\Delta_\odot$ to $J_2$ we use Eq.~\eqref{eq:DickeJ2plus},  including only the $J_2$ contribution and assuming $\Delta r_{\rm surf} = 7.78\ \rm mas$~\cite{Dicke1970oblatenessJ2ApJ}. In the Reference column, we indicate the first author followed by the date(s) of observation of the corresponding measurement.}
    \end{threeparttable}
\end{table*}

\setlength{\tabcolsep}{5pt}
\renewcommand{\arraystretch}{1.3}
    
\begin{table}[!htbp]
        \begin{minipage}[t]{.5\textwidth}
      \centering
        \begin{threeparttable}
    \label{tab:summary_J2_helio}
    \begin{tabularx}{0.95\textwidth}{@{} l S[table-format=-3.2] S[table-format=-3.2] c c @{}}
    \toprule
    \multicolumn{5}{c}{Helioseismological $J_2$ measurements} \\ \toprule
    \# & {$J_2 (\times 10^{-7})$} & {$\pm J_2 (\times 10^{-7})$} & Reference & Solar Model \\
    \midrule
    1 & 36 & {--} & Gough~\cite{Gough1982InternalRA} & \cite{Christensen1979SolarModel, Christensen1979SolarModelErratum} \\
    2 & 1.7 & 0.4 & Duvall~\cite{Duvall1984Helio} &  \cite{Christensen1979SolarModel, Christensen1979SolarModelErratum} \\
    3 & 55 & 13 & Hill~\cite{Hill1986Helio}\tnote{ a}& --\tnote{b}\\ \cdashline{1-5}[0.1pt/0.5pt] 
    4 & 2.23 & 0.09 & Pijpers~\cite{Pijpers1998helioJ2}& \cite{Christensen1996SolarModelS}\\
    5 & 2.22 & 0.02 & Armstrong~\cite{Armstrong1999heliomeasurement} &\cite{Christensen1992SolarModel}\\
    6 & 1.6 & 0.04 & Godier~\cite{Godier1999HelioJ2, Godier1999heliomeasurement}\tnote{c}& \cite{Richard1996SolarModel}\\
    7 & 2.201 & {--} & Mecheri~\cite{Mecheri2004J2helio}&\cite{Corbard2002SolarModel}\tnote{d}\\
    8 & 2.198 & {--} & Mecheri~\cite{Mecheri2004J2helio}& \cite{Corbard2002SolarModel}\\
    9 & 2.220 & 0.009 & Antia~\cite{Antia2008Helio}\tnote{e}&\\
    10 & 2.204 & {--} & Mecheri~\cite{Mecheri2021J2helio}&\cite{CESAM2008}\\
    11 & 2.208 & {--} & Mecheri~\cite{Mecheri2021J2helio}&\cite{ASTEC2008}\\ \cdashline{1-5}[0.1pt/0.5pt] 
    12 & 2.206 & {--} & Roxburgh~\cite{Roxburgh2001heliomeasurement}&\cite{Christensen1996SolarModelS}\\ 
    13 & 2.208 & {--} & Roxburgh~\cite{Roxburgh2001heliomeasurement}&\cite{Marchenkov1998rotationalmodel}\\ \cdashline{1-5}[0.1pt/0.5pt] 
    14 & 2.14 & 0.09 & Pijpers~\cite{Pijpers1998helioJ2}& \cite{Christensen1996SolarModelS}\\
    15 & 2.18 & {--} & Antia~\cite{Antia2000heliomeasurement}&\\
    16 & 2.180 & 0.005  & Antia~\cite{Antia2008Helio}\tnote{e}&\\ \cdashline{1-5}[0.1pt/0.5pt] 
    17 & 2.211 & {--} & Mecheri~\cite{Mecheri2021J2helio}& \cite{CESAM2008} \\
    18 & 2.216 & {--} & Mecheri~\cite{Mecheri2021J2helio}&\cite{ASTEC2008}\\
    \bottomrule
    \end{tabularx}
    \begin{tablenotes}
        \small
        \item[a] This is the corrected value after accounting for a missing numerical factor~\cite{Rozelot2011historyoblateness}.
        \item[b] Solar model by Saio, H. (1982), private communication~\cite{Hill1986Helio}.
        \item[c] The uncertainty is reported in \cite{Zwaard2022summarytablemeasurements}.
        \item[d] Measurements 7 and 8 use Model (a) and (b) for the solar
        differential rotation from \cite{Corbard2002SolarModel}, respectively. 
        \item[e] The central value and error 
        come from the \textit{time average} of $J_2$ measurements over 10 years.
    \end{tablenotes}
    \caption{\label{tab:J2:helio}Summary of helioseismological $J_2$ measurements. 
    Measurements \#1--3 use  the rotational splitting data from~\cite{Hill1985Rotsplitting}, \#4--11 use data from SoHO/MDI~\cite{Scherrer1995SOHOMDImeasurement, Larson2015helioanal}, 
    \#12--13 use data from SoHO/MDI and the ground-based Global Oscillation Network Group (GONG)~\cite{Hill1996GONGmeasurement}, 
    \#14--16 use data from GONG, 
    and 
    \#17--18 use data from SDO/HMI~\cite{Larson2018helioanal}. }
    \end{threeparttable}
    \end{minipage} 
    \begin{minipage}[t]{.4\textwidth}
          \centering
           \begin{threeparttable}
    \label{tab:summary_J2_Orbital}
    \begin{tabularx}{\textwidth}{@{} l  S[table-format=-3.2] S[table-format=-3.2] c @{}}
    \toprule
    \multicolumn{4}{c}{Orbital $J_2$ measurements} \\ \toprule
     \# & {$J_2 (\times 10^{-7})$} & {$\pm J_2 (\times 10^{-7})$} & Reference \\
    \midrule
    1 & 2.25 & 0.09 & Park ,11--14~\cite{Park2017MESSENGER}\\
    2 & 2.246 & 0.022 & Genova, 08--15~\cite{Genova2018MESSENGER}\\  \cdashline{1-4}[0.1pt/0.5pt] 
    3 & 2.165 & 0.12 &Fienga, -~\cite{Fienga2019eph}\\
    4 & 2.206 & 0.03 &Fienga, -~\cite{Fienga2019eph}\\
    \bottomrule
    \end{tabularx}
     \caption{\label{tab:J2:orb}Summary of orbital $J_2$ measurements. Measurements \#1 and \#2 are orbital assessments using data almost exclusively 
     from measurements of Mercury's orbit. We also include the ephemerides  
     analyses 
     that have taken into account 
     the ELT effect and fit for 
     $J_2$ together with PPN parameters.}
    \end{threeparttable}
    \end{minipage}%
\end{table}

\end{widetext}

\setlength{\tabcolsep}{5pt}
\renewcommand{\arraystretch}{1.3}

\begin{table}[!htbp]
 \begin{threeparttable}
    \begin{tabularx}{0.5\textwidth}{@{} l  S[table-format=-3.2] S[table-format=-3.2] c c  c @{}}
    \toprule
    \multicolumn{6}{c}{Orbital $J_2$ measurements} \\ \toprule
     \# & {$J_2 (\times 10^{-7})$} & {$\pm J_2 (\times 10^{-7})$} & ELT & PPN & Reference \\
    \midrule
    1 & 180 &  200 & N & N & Lieske, 49--68~\cite{Lieske1969ICARUS}\\
     2 & 13.9 &  24.7 & N & Y & Shapiro, 66--71~\cite{Shapiro1972Orb}\\
     3 & 25 &  16 & N & Y & Anderson, 11--76~\cite{Anderson1976orb}\\
      4 & 12.3 &  11.5 & N & N & Anderson, --~\cite{Pitjeva2014Ephemerides}\\
     5 & -1.8 &  4.5 & N & Y & Eubanks, --~\cite{Pitjeva2014Ephemerides}\\
     6 & -11.7 & 9.5 & N & Y &  Pitjeva, --~\cite{Pitjeva2014Ephemerides} \\
     7 & -1.3 & 4.1 & N & Y &  Pitjeva, 64-89~\cite{Pitjeva1993orbassess} \\
     8 & 2.4 & 0.7 & N & Y &  Pitjeva, --~\cite{Pitjeva2001Eph} \\
    9 & -5 & 10 & N & Y &  Williams, 96--00~\cite{Williams2002orb} \\
    10 & 6.6 & 9.0&  N & N & Afanaseva, 80--86~\cite{Afanaseva1990Orb}\\
    11 & -6 & 58& N & N & Landgraf, 49--87~\cite{Landgraf1992Orb}\\
    12 & 2.3 & 5.2&  N & Y & Anderson, 71--97~\cite{Anderson2002orb}\\
    13 & 1.9 & 0.3 & N & Y &  Pitjeva, 61--03~\cite{Pitjeva2005OblatenessOrb} \\
    14 & 2.22 & 0.23 & N & Y & Pitjeva, --~\cite{Pitjeva2014Ephemerides}\\
    15& 2.25 & 0.09 & Y & Y & Park ,11--14~\cite{Park2017MESSENGER}\\
    16 & 2.246 & 0.022 & Y & Y & Genova, 08--15~\cite{Genova2018MESSENGER}\\ 
    17 & 2.46 & 0.68 & N & Y& Standish, -~\cite{Standish2006orb}\\
    18 & 2.295 & 0.010 & N & N & Viswanathan, -~\cite{Viswanathan2017Eph}\\
     19 & 1.82 & 0.47 & N & N & Fienga, -~\cite{Fienga2009ephemerides}\\
    20 & 1.8 & ? & N & Y & Konopliv,--~\cite{Konopliv2011Orb}\\
    21 & 2.0 & 0.20 & N & Y & Pitjeva,--~\cite{Pitjeva2013releffectsdm}\\
    22 & 2.40 & 0.25& N & Y &Fienga, -~\cite{Fienga2011ephmerides}\\
    23 & 2.27 & 0.25& N & Y &Fienga, -~\cite{2015fiengaINPOP}\\
    24 & 2.22 & 0.13& N & Y &Fienga, -~\cite{2015fiengaINPOP}\\
    25 & 2.165 & 0.12& Y & Y &Fienga, -~\cite{Fienga2019eph}\\
    26 & 2.206 & 0.03& Y & Y &Fienga, -~\cite{Fienga2019eph}\\
    27 & 2.40 & 0.20 & N & Y &  Verma, -~\cite{Verma2014Messenger}\\
    28 & 2.010 & 0.010 & N & N & Fienga, -~\cite{Fienga2019OEphemerides}\\
    29 & 2.2180 & 0.01 & Y & N & Fienga, -~\cite{Fienga2021Ephemerides_INPOP21a}\\
    \bottomrule
    \end{tabularx}
     \caption{Extended summary of the orbital $J_2$ assessments including 
     information as to whether the references included a simultaneous fit of the PPN parameters or the ELT effect correction in their analysis.}
    \label{tab:J2_Orb_extended}\end{threeparttable}
\end{table}

\section{Dark Halo Formation}
\label{sec:app:graviatom}
We follow \cite{Budker2023gravatom} to 
find that 
a gravi-atom radius is set by 
\begin{equation}
R_\ast  = \frac{(\hbar c)^2 c^2}{GM (mc^2)^2} \approx 0.035\, {\rm AU}
\left( \frac{10^{12}\, {\rm kg}}{M} \right) \left( \frac{100\, \mu{\rm eV}}{m} \right)^2 \,, 
\end{equation} 
noting that $0.035\, {\rm AU}\approx 7.5\, R_{\odot}$. 
The gravitational coupling 
$\alpha$ is determined by 
$R_{\ast}= (m\alpha)^{-1}$, 
and the bound-state 
escape velocity is 
$\sqrt{2} \alpha$.  
The associated de Broglie wavelength $\lambda_{\rm db}$ 
of a dark-matter particle is 
\begin{equation}
\lambda_{\rm db} \approx 1.6 \times 10^{1}\, {\rm m} 
 \left( \frac{100\, \mu {\rm eV}}{m} \right) \left( \frac{240\, {\rm km/s}}{v} \right) \,, 
\end{equation}
where $v$ is the particle speed. 
Dark matter is wave-like if $\lambda_{\rm db}$ exceeds
the dark inter-particle spacing, possible if $m\lesssim 30 \rm eV$. 
Moreover, 
\begin{equation}
    \xi_{\rm foc} \approx 2.9 \times 10^{-9} \left( \frac{M}{10^{12}\, {\rm kg}} \right) \left( \frac{m}{100\, \mu{\rm eV}} \right)  \left( \frac{240\, {\rm km/s}}{v} \right)  
\end{equation}
If $\xi_{\rm foc} \gtrsim 1$, the expected maximum overdensity, with 
$\delta \rho_{\rm dm} \sim \rho_{\rm dm}$, is 
\begin{equation}
\delta\rho_{\rm dm} = 7 \times 10^{13} 
\left( \frac{f_a}{10^{11} {\rm GeV}} \right)^2 
\left( \frac{M}{10^{12} {\rm kg}} \right)^2 \left( \frac{m}{100 \mu{\rm eV}} \right)^4  
\,, 
\end{equation}
which can be reached over a time scale of about
\begin{equation}
\tau_{\rm rel} \approx 9 \times 10^{12}\, {\rm Gyr} 
\left( \frac{f_a}{10^{5} {\rm GeV}} \right)^4
\left( \frac{m}{1 \mu{\rm eV}} \right)^3  
\left( \frac{240 {\rm km/s}}{v} \right)^2
\,,
\end{equation}
where $\tau_{\rm rel} \sim \rho_{\rm dm}^{-2}$. 
If $\lambda > 0$, then the ultimately formed halo should be stable.

\bibliography{Sun-withDM_arxiv-v2_PRD_final_8Apr25.bbl}

\end{document}